\documentclass[iop]{emulateapj}
\usepackage{color} 

\usepackage{ulem}
\usepackage{amsmath,apjfonts}
\usepackage{enumerate}

\newcommand\ergs{erg~s$^{-1}$}

\newcommand\Var{\mathop{\rm Var}} 
\newcommand\Cov{\mathop{\rm Cov}} 
\newcommand{\bracket}[1]{\left\langle#1\right\rangle}

\newcommand{\hbeta}{H{$\beta$}}
\newcommand{\hbetawave}{H{$\beta$}\,$\lambda$4861}
\newcommand{\halpha}{H{$\alpha$}}

\newcommand{\CIV}{C\,{\sc iv}}
\newcommand{\CIVwave}{C\,{\sc iv}\,$\lambda$1549}
\newcommand{\MgII}{Mg\,{\sc ii}}
\newcommand{\MgIIwave}{Mg\,{\sc ii}\,$\lambda$2798}

\newcommand{\OIII}{[O\,{\sc iii}]}

\newcommand{\OIIIb}{[O\,{\sc iii}]\,$\lambda$5007}
\newcommand{\OIIIab}{[O\,{\sc iii}]\,$\lambda\lambda$4959,5007}
\newcommand{\FeII}{Fe\,{\sc ii}}

\newcommand{\HeIIwave}{He\,{\sc ii}\,$\lambda$4686}

\begin{document}

\title{Single-Epoch Black Hole Mass Estimators For Broad-Line Active Galactic Nuclei: Recalibrating \hbeta\ with A New Approach}

\author{Hua Feng\altaffilmark{1}, Yue Shen\altaffilmark{2,3}, and Hong Li\altaffilmark{1}}

\altaffiltext{1}{Department of Engineering Physics and Center for Astrophysics, Tsinghua University, Beijing 100084, China}
\altaffiltext{2}{Carnegie Observatories, 813 Santa Barbara Street, Pasadena, CA 91101, USA}
\altaffiltext{3}{Hubble Fellow}

\shorttitle{Empirical Scaling Between Mass and Luminosities in AGN}
\shortauthors{Feng, Shen, \& Li}

\begin{abstract}

Based on an updated \hbeta\ reverberation mapping (RM) sample of 44 nearby active galactic nuclei (AGN), we propose a novel approach for black hole (BH) mass estimation using two filtered luminosities computed from single-epoch (SE) AGN spectra around the \hbeta\ region. We found that the two optimal-filter luminosities extract virial information (size and virial velocity of the broad line region, BLR) from the spectra, justifying their usage in this empirical BH mass estimator. The major advantages of this new recipe over traditional SE BH mass estimators utilizing continuum luminosity and broad line width are: 1) it has a smaller intrinsic scatter of 0.28~dex calibrated against RM masses; 2) it is extremely simple to use in practice, without any need to decompose the spectrum; 3) it produces unambiguous and highly repeatable results even with low signal-to-noise spectra. The combination of the two luminosities can also cancel out, to some extent, systematic luminosity errors potentially introduced by uncertainties in distance or flux calibration. In addition, we recalibrated the traditional SE mass estimators using broad \hbeta\ full width at half maximum (FWHM) and monochromatic continuum luminosity at 5100\AA\ ($L_{\rm 5100}$). We found that using the best-fit slopes on FWHM and $L_{\rm 5100}$ (derived from fitting the BLR radius-luminosity relation and the correlation between rms line dispersion and SE FWHM, respectively) rather than simple assumptions (e.g., 0.5 for $L_{\rm 5100}$ and 2 for FWHM) leads to more precise SE mass estimates, improving the intrinsic scatter from 0.41~dex to 0.36~dex with respect to the RM masses. We compared different estimators and discussed their applications to the Sloan Digital Sky Survey (SDSS) quasar sample.  Due to the limitations of the current RM sample, application of any SE recipe calibrated against RM masses to distant quasars should be treated with caution.
\end{abstract}


\section{Introduction}
\label{sec:intro}

Active galactic nuclei (AGN) are known to be powered by accretion onto central supermassive black holes in galaxies. Accurate estimation of black hole masses in AGN is of great interest in astrophysics, not only because mass is one of the two fundamental parameters of an astrophysical black hole and is directly related to accretion physics, but also because it plays a key role in understanding the cosmological growth and evolution of supermassive black holes along with their host galaxies \citep[][and references therein]{Ferrarese2000,Gebhardt2000,Kormendy_Ho_2013}. 

Based on the assumption that the broad line region (BLR) in AGN is virialized \citep[e.g.,][]{Peterson1999,Peterson2000}, the reverberation mapping (RM) method \citep[e.g.,][]{Blandford_McKee_1982,Peterson_1993} has been widely applied to measure the central black hole mass in broad-line AGN following the equation
\begin{equation}
M_{\rm RM} = f \frac{R \Delta V^2}{G},
\end{equation}
where $R$ is the average BLR radius, $\Delta V$ is the broad line width (as a proxy for the BLR virial velocity), $G$ is the gravitational constant, and $f$ is the virial coefficient that accounts for the geometry and kinematics of the BLR \citep[for a review see, e.g.,][]{Peterson2011,Shen2013}. In practice, the BLR size is measured from RM via the time lag ($\tau$) between the broad line and continuum variabilities ($R = c\tau$), and the line width is usually characterized by the broad line dispersion (second moment, $\sigma_{\rm rms}$) measured from the root-mean-square (rms) spectrum during the RM campaign. 

RM studies of nearby ($z<0.3$) AGN have led to the discovery of a radius-luminosity ($R-L$) relation for BLRs \citep[e.g.,][]{Kaspi2000,Kaspi2005,Bentz2009a,Bentz2013},
\begin{equation}
R \propto L^\alpha,
\end{equation}
where the slope $\alpha$ was found to be close to 0.5, as expected in simple photoionization theory, when contamination from host galaxy starlight is properly removed in estimating the AGN luminosity $L$ \citep{Bentz2009a,Bentz2013}. This allows one to substitute simple luminosity measurements for expensive time-lag measurements, and allows for black hole mass estimates with single-epoch (SE) spectra, 
\begin{equation}\label{eqn:SE}
\log(M_{\rm SE}) = \alpha \log(L) + \beta \log(\Delta V_{\rm SE}) + {\rm constant},
\end{equation}
where $L$ and $\Delta V_{\rm SE}$ are measured from SE spectra around different broad lines, and the coefficients are calibrated using the RM BH masses in the local RM AGN sample as standard. These SE estimators make it possible to estimate virial BH masses for a large number of AGN including high redshift quasars with SE spectroscopy \citep[e.g.][]{Vestergaard2006}.

However, there are a number of caveats one should be aware of with SE virial mass estimators \citep[for a review see, e.g.,][]{Shen2013}. First of all, the broad line width $\Delta V$ for RM is measured from the rms spectrum to minimize contamination by non-reverberating components, while $\Delta V_{\rm SE}$ in Equation\ (\ref{eqn:SE}) is measured from SE spectra. A correlation between $\Delta V$ and $\Delta V_{\rm SE}$ is the basic requirement to justify the usage of $\Delta V_{\rm SE}$ in SE virial mass estimators. Different definitions of line width are proposed, such as the full width at half maximum (FWHM) or the line dispersion \citep[the second moment of the line;][]{Peterson2004}. Both definitions have their own advantages and disadvantages. Arguably line dispersion is a better proxy for the underlying virial velocity than FWHM, although the evidence is circumstantial \cite[see discussion in, e.g.,][]{Shen2013}. Measuring FWHM becomes particularly difficult and may lead to biased results if the line profile is complex or the narrow line component cannot be removed properly. On the other hand, line dispersion measurements are sensitive to the specific treatment of line blending, and are more susceptible to low signal-to-noise ratio ($S/N$) than FWHM, where the wing fluxes can be lost in noise.  

Various broad emission lines have been advocated as SE virial mass estimators, notably the Balmer lines (\hbeta\ and \halpha), \MgII, and \CIV, calibrated either against RM BH masses (for the local RM sample), or internally among different SE estimators with large statistical quasar samples \citep[e.g.,][and references therein]{Shen2013}. \hbeta\ is the most widely used line, as this line is the best studied broad line in RM, and the current $R-L$ relation is only {\it robustly} established for the \hbeta\ BLR. \MgIIwave\ is usually adopted for objects at intermediate redshifts and shows consistent results with that of \hbeta\ \citep{McLure2004,McGill_etal_2008,Shen_etal_2008,Shen2012}, but little RM data exist for \MgII. For quasars at even higher redshift, \CIVwave\ is the primary choice. The \CIV\ time lags were measured from several objects in the \hbeta\ RM catalog \citep{Peterson2004} plus an intermediate mass black hole \citep{Peterson2005} and a high luminosity quasar at $z = 2.17$ \citep{Kaspi2007}. However, mounting evidence suggests that \CIV\ may be a biased virial BH mass estimator, especially at high luminosities \citep[e.g.,][]{Baskin_Laor_2005,Sulentic_etal_2007,Netzer_etal_2007,Shen_etal_2008,Denney2012}. Its broadening is perhaps a combination of both virial motion and outflows \citep[e.g.,][and references therein]{Shen_etal_2008,Richards2011,Wang_etal_2011}, and additional correction may be needed to reduce the non-virial contribution in the \CIV\ line width \citep[e.g.,][]{Denney2012}. A test of the three lines using repeated quasar spectra in the Sloan Digital Sky Survey (SDSS) DR7 \citep{Abazajian2009} suggests that \hbeta\ may be the only one that varies in a manner consistent with the virial assumption: the line widths of \MgII\ and \CIV\ fail to respond to the variation of continuum luminosities, rather they seem to be constant despite the change of luminosity \citep[][]{Shen2013}. 

While it is popular to use a calibrated SE virial mass estimator to estimate AGN BH masses with SE spectra, one needs to follow the specific definitions of these recipes and proper procedures to measure spectral properties to derive unbiased (relative to the specific calibration) BH mass estimates. Given the variations in these SE mass recipes and their different dependences on spectral quality, different groups often derive different results even for the same set of data. This drives the need for developing empirical SE mass estimators that are not only calibrated to be consistent with the reference BH masses, but also are robust to the details of spectral measurement procedures and spectral quality when applied to other objects, which is the main motivation of this work.

Here we re-calibrate the SE virial mass estimators based on \hbeta, using an updated sample of local RM AGN. We choose \hbeta\ because the observed $R-L$ relation is directly based on this line, and most of the RM AGN have spectroscopic coverage of this line, yielding the largest calibration sample with RM masses. As argued above, \hbeta\ is also the most secure line to use as a virial mass estimator. Our main goal is to design objective SE mass recipes that are easy and robust to use, and offer considerably better stability on low-spectral quality data compared with earlier SE recipes. 

The paper is organized as follows. In \S\ref{sec:sam} we describe the sample construction and spectral measurements. Recalibration of traditional SE virial mass estimators is described in \S\ref{sec:recal}. A new SE mass estimator based on filtered luminosities is presented in \S\ref{sec:opt}. The different estimators are tested and compared in \S\ref{sec:test} and the results are discussed in \S\ref{sec:dis}.

\section{Sample}
\label{sec:sam}

Our sample includes local AGN with published RM data. A list of 44 low-redshift AGN is compiled from several major RM campaigns: \citet{Peterson2004}, \citet{Denney2010}, \citet{Grier2012}, and the Lick AGN Monitoring Program \citep{Bentz2009c,Barth2011a,Barth2011b}. For objects included in different campaigns, the latest results are adopted, except for Mrk 817 whose rms spectrum from the new measurement was found to be problematic \citep{Denney2010}, so the previous measurement from \citet{Peterson2004} is adopted. The RM masses for some objects, including PG 1211+143, NGC 4593, and IC 4329A, are claimed to be unreliable by the authors and are not used \citep{Peterson2004}. The sample of \citet{Peterson2004} relies on measurements of multiple emission lines rather than \hbeta\ only. We hence choose only \hbeta\ observations. If there are more than one \hbeta\ observations, we take the line width and time delay from the one whose virial product has the smallest relative error; for the RM mass, we use all \hbeta\ results. This further removes one object, PG 0844+349, because the time delay was not measured for \hbeta\ \citep{Peterson2004}. 

The object names and related properties are listed in Table~\ref{tab:src}, including the rest-frame time delay $\tau_{\rm cent}$ (centroid of the cross correlation), the line widths in line dispersion ($\sigma_{\rm rms}$) measured from the rms spectra, and the RM mass defined as $M_{\rm RM} = fc\tau_{\rm cent} \sigma^2_{\rm rms}/ G$. If multiple measurements are available \citep[for objects in][]{Peterson2004}, the RM mass is a weighted mean of individual virial products times the virial coefficient. To be consistent, the masses for Zw 229-015 and Mrk 50 \citep{Barth2011a,Barth2011b} are rescaled using a virial coefficient $f = 5.5$ \citep{Onken2004} as for all other objects. We do not use new $f$ values calibrated more recently \citep[e.g.,][]{Woo2010,Graham_etal_2011} because the determination of the average $f$ value is still subject to significant uncertainties, and we note that our results can be simply scaled by a constant when more accurate $f$ values become available in the future\footnote{We note here $f$ is the average virial coefficient that does not account for the diversity in individual BLRs and orientation effects.}. For objects with SE spectra available from \citet{Marziani2003}, their redshifts $z$ are also adopted. Otherwise, the redshift is obtained from the NASA/IPAC Extragalactic Database (NED) except for six objects whose redshifts in NED are inconsistent with data by more than 0.0002; new redshifts are estimated for them based on their spectra and listed in Table~\ref{tab:src}. The luminosity distance $d_L$ is calculated from redshift assuming a standard cosmology with $h = 0.7$, $\Omega_m = 0.3$, and $\Omega_\Lambda = 0.7$, except for two cases at $z < 0.005$ where independent distance measurements are available (see Table~\ref{tab:src} for references). 

\begin{deluxetable*}{cllllllcc}
\tablecolumns{9}
\tablewidth{0pc}
\tabletypesize{\scriptsize}
\tablecaption{An \hbeta-based reverberation mapping sample \label{tab:src}}

\tablehead{
\colhead{no.} & \colhead{name} & \colhead{$\tau_{\rm cent}$} & \colhead{$\sigma_{\rm rms}$} & \colhead{$M_{\rm RM}$} & \colhead{ref.} & 
\colhead{$z$} & \colhead{$d_L$} & \colhead{$E(B-V)$} \\
\colhead{} & \colhead{} & \colhead{(days)} & \colhead{(km s$^{-1}$)} & \colhead{($10^6 M_\sun$)} & \colhead{} & \colhead{} & \colhead{(Mpc)} & \colhead{} \\
\colhead{(1)} & \colhead{(2)} & \colhead{(3)} & \colhead{(4)} & \colhead{(5)} & \colhead{(6)} & \colhead{(7)} & \colhead{(8)} & \colhead{(9)} 
}
\startdata
01 & PG 0026+129 & $111.0^{+24.1}_{-28.3}$ & $1773\pm285$ & $375^{+145}_{-154}$ & P04 & 0.1420 & 672 & 0.071 \\
02 & PG 0052+251 & $89.8^{+24.5}_{-24.1}$ & $1783\pm86$ & $306^{+89}_{-87}$ & P04 & 0.1550 & 739 & 0.047 \\
03 & Fairall 9 & $17.4^{+3.2}_{-4.3}$ & $3787\pm197$ & $268^{+57}_{-72}$ & P04 & 0.0460 & 204 & 0.026 \\
04 & Mrk 590 & $29.2^{+4.9}_{-5.0}$ & $1251\pm72$ & $47.5\pm7.4$ & P04 & 0.0270 & 118 & 0.037 \\
05 & Akn 120 & $37.1^{+4.8}_{-5.4}$ & $1884\pm48$ & $150\pm19$ & P04 & 0.0330 & 145 & 0.127 \\
06 & Mrk 79 & $16.0^{+6.4}_{-5.8}$ & $1854\pm72$ & $52.4\pm14.3$ & P04 & 0.0222 & 96.7 & 0.071 \\
07 & PG 0804+761 & $146.9^{+18.8}_{-18.9}$ & $1971\pm105$ & $613\pm102$ & P04 & 0.1000 & 460 & 0.035 \\
08 & Mrk 110 & $24.3^{+5.5}_{-8.3}$ & $1196\pm141$ & $25.1\pm5.9$ & P04 & 0.0360 & 158 & 0.013 \\
09 & PG 0953+414 & $150.1^{+21.6}_{-22.6}$ & $1306\pm144$ & $275^{+72}_{-73}$ & P04 & 0.2347 & 1173 & 0.012 \\
10 & NGC 3783 & $10.2^{+3.3}_{-2.3}$ & $1753\pm141$ & $33.8^{+12.3}_{-9.3}$ & P04 & 0.0090 & 38.8 & 0.121 \\
11 & NGC 4151 & $3.1\pm1.3$ & $1914\pm42$ & $12.0\pm5.0$ & P04 & 0.0030 & 12.9 & 0.027 \\
12 & PG 1226+023 & $306.8^{+68.5}_{-90.9}$ & $1777\pm150$ & $1040^{+292}_{-352}$ & P04 & 0.1580 & 755 & 0.021 \\
13 & PG 1229+204 & $37.8^{+27.6}_{-15.3}$ & $1385\pm111$ & $78.1^{+58.3}_{-34.1}$ & P04 & 0.0640 & 287 & 0.027 \\
14 & PG 1307+085 & $105.6^{+36.0}_{-46.6}$ & $1820\pm122$ & $376^{+138}_{-173}$ & P04 & 0.1550 & 739 & 0.035 \\
15 & Mrk 279 & $16.7\pm3.9$ & $1420\pm96$ & $36.3\pm9.8$ & P04 & 0.0310 & 136 & 0.016 \\
16 & PG 1411+442 & $124.3^{+61.0}_{-61.7}$ & $1607\pm169$ & $344^{+184}_{-186}$ & P04 & 0.0890 & 407 & 0.009 \\
17 & PG 1426+015 & $95.0^{+29.9}_{-37.1}$ & $3442\pm308$ & $1210^{+440}_{-517}$ & P04 & 0.0860 & 392 & 0.032 \\
18 & Mrk 817 & $19.0^{+3.9}_{-3.7}$ & $1392\pm78$ & $49.4\pm7.7$ & P04 & 0.0330 & 145 & 0.007 \\
19 & PG 1613+658 & $40.1^{+15.0}_{-15.2}$ & $2547\pm342$ & $279\pm129$ & P04 & 0.1290 & 605 & 0.027 \\
20 & PG 1617+175 & $71.5^{+29.6}_{-33.7}$ & $2626\pm211$ & $529^{+235}_{-263}$ & P04 & 0.1140 & 530 & 0.042 \\
21 & PG 1700+518 & $251.8^{+45.9}_{-38.8}$ & $1700\pm123$ & $781^{+182}_{-165}$ & P04 & 0.2900 & 1493 & 0.034 \\
22 & 3C 390.3 & $23.6^{+6.2}_{-6.7}$ & $3105\pm81$ & $244^{+66}_{-71}$ & P04 & 0.0570 & 255 & 0.072 \\
23 & Mrk 509 & $79.6^{+6.1}_{-5.4}$ & $1276\pm28$ & $139\pm12$ & P04 & 0.0350 & 154 & 0.057 \\
24 & NGC 7469 & $4.5^{+0.7}_{-0.8}$ & $1456\pm207$ & $10.1^{+3.2}_{-3.4}$ & P04 & 0.0170 & 73.8 & 0.069 \\
25 & Mrk 142 & $2.75^{+0.58}_{-0.63}$ & $859\pm102$ & $2.17^{+0.77}_{-0.83}$ & B09 & 0.0449 & 199 & 0.017 \\
26 & SBS 1116+583A & $2.25^{+0.44}_{-0.38}$ & $1528\pm184$ & $5.80^{+2.09}_{-1.86}$ & B09 & 0.0279 & 122 & 0.012 \\
27 & Arp 151 & $3.79^{+0.42}_{-0.49}$ & $1252\pm46$ & $6.72^{+0.96}_{-1.24}$ & B09 & 0.0207$^\dagger$ & 90.0 & 0.014 \\
28 & Mrk 1310 & $3.62^{+0.36}_{-0.38}$ & $755\pm138$ & $2.24\pm0.90$ & B09 & 0.0196 & 85.0 & 0.030 \\
29 & Mrk 202 & $3.05^{+0.79}_{-0.78}$ & $659\pm65$ & $1.42^{+0.85}_{-0.59}$ & B09 & 0.0231$^\dagger$ & 101 & 0.020 \\
30 & NGC 4253 & $6.45^{+0.97}_{-1.01}$ & $516\pm218$ & $1.76^{+1.56}_{-1.40}$ & B09 & 0.0129 & 55.9 & 0.020 \\
31 & NGC 4748 & $5.99\pm1.21$ & $657\pm91$ & $2.57^{+1.03}_{-1.25}$ & B09 & 0.0139$^\dagger$ & 60.0 & 0.051 \\
32 & NGC 6814 & $6.56^{+0.64}_{-0.66}$ & $1610\pm108$ & $18.5\pm3.5$ & B09 & 0.0052 & 22.4 & 0.185 \\
33 & Mrk 290 & $8.72^{+1.21}_{-1.02}$ & $1609\pm47$ & $24.3\pm3.7$ & D10 & 0.0300 & 131 & 0.014 \\
34 & NGC 3227 & $3.75^{+0.76}_{-0.82}$ & $1376\pm44$ & $7.63^{+1.62}_{-1.72}$ & D10 & 0.0039 & 23.6$^\ddagger$ & 0.023 \\
35 & NGC 3516 & $11.68^{+1.02}_{-1.53}$ & $1591\pm10$ & $31.7^{+2.8}_{-4.2}$ & D10 & 0.0088 & 38.1 & 0.042 \\
36 & NGC 4051 & $1.87^{+0.54}_{-0.50}$ & $927\pm64$ & $1.73^{+0.55}_{-0.52}$ & D10 & 0.0022$^\dagger$ & 14.5$^\ddagger$ & 0.013 \\
37 & NGC 5548 & $12.40^{+2.74}_{-3.85}$ & $1822\pm35$ & $44.2^{+9.9}_{-13.8}$ & D10 & 0.0167$^\dagger$ & 72.3 & 0.020 \\
38 & Mrk 335 & $14.3\pm0.7$ & $1293\pm64$ & $25\pm3$ & G12 & 0.0250 & 109 & 0.035 \\
39 & Mrk 1501 & $12.6\pm3.9$ & $3321\pm107$ & $184\pm27$ & G12 & 0.0900 & 411 & 0.099 \\
40 & 3C 120 & $25.9\pm2.3$ & $1514\pm65$ & $67\pm6$ & G12 & 0.0330 & 145 & 0.297 \\
41 & Mrk 6 & $10.1\pm1.1$ & $3714\pm68$ & $136\pm12$ & G12 & 0.0188 & 81.7 & 0.136 \\
42 & PG 2130+099 & $9.6\pm1.2$ & $1825\pm65$ & $46\pm4$ & G12 & 0.0630 & 283 & 0.044 \\
43 & Zw 229-015 & $3.86^{+0.69}_{-0.90}$ & $1590\pm47$ & $10.51^{+1.98}_{-2.53}$ & B11a & 0.0275$^\dagger$ & 120 & 0.072 \\
44 & Mrk 50 & $10.40^{+0.80}_{-0.91}$ & $1740\pm101$ & $34.1\pm5.0$ & B11b & 0.0234 & 102 & 0.016 \\
\enddata

\tablerefs{
References for RM information: 
B11a -- \citet{Barth2011a};
B11b -- \citet{Barth2011b};
B09 -- \citet{Bentz2009c};
D10 -- \citet{Denney2010};
G12 -- \citet{Grier2012};
P04 -- \citet{Peterson2004}.}

\tablenotetext{$\dagger$}{Redshifts are adopted from SDSS DR7
\citep{Abazajian2009} for Arp 151 and Mrk 202, 
\citet{Jones2009} for NGC 4748, and 
\citet{Falco1999} for NGC 4051, NGC 5548, and Zw 229-015.} 

\tablenotetext{$\ddagger$}{Independent luminosity distances are adopted from
\citet{Russell2002} for NGC 4051 with the mean Tully-Fisher distance, and
\citet{Blakeslee2001} for NGC 3227 with the surface brightness fluctuation distance to its interacting companion NGC 3226.}

\end{deluxetable*}

\begin{deluxetable*}{clllcclc}
\tablecolumns{8}
\tablewidth{0pc}
\tabletypesize{\scriptsize}
\tablecaption{Data source and spectral information \label{tab:spec}}

\tablehead{
\colhead{no.} & \colhead{name} & \colhead{source} & \colhead{obs.\ date} & \colhead{FWHM$_{\rm inst}$} & \colhead{$\log[L_{5100}/({\rm erg\,s^{-1}})]$} & \colhead{FWHM} & \colhead{$\log[L_{\rm Bulge}/({\rm erg\,s^{-1}})]$}\\
\colhead{} & \colhead{} & \colhead{} & \colhead{} & \colhead{(\AA)} & \colhead{} & \colhead{(km s$^{-1}$)} & \colhead{}\\
\colhead{(1)} & \colhead{(2)} & \colhead{(3)} & \colhead{(4)} & \colhead{(5)} & \colhead{(6)} & \colhead{(7)} & \colhead{(8)}
}
\startdata
01 & PG 0026+129 & M03 & 1990-10-11 & 6.5 & 44.70 & $2598\pm57$ & 44.36\\
02 & PG 0052+251 & M03 & 1994-12-09 & 7 & 44.77 & $5318\pm118$ & 44.12\\
03 & Fairall 9 & M03 & 1993-12-20 & 4.2 & 43.90 & $5618\pm107$ & 44.10\\
04 & Mrk 590 & M03 & 1996-10-13 & 3 & 44.07 & $2966\pm56$ & 43.38\\
05 & Akn 120 & M03 & 1990-04-03 & 3.5 & 44.42 & $5987\pm54$ & 44.01\\
06 & Mrk 79 & P98 & 1989-11-06 & 10 & 43.68 & $4735\pm44$ & 43.14\\
07 & PG 0804+761 & M03 & 1990-02-16 & 6.5 & 44.61 & $3190\pm39$ & 44.18\\
08 & Mrk 110 & M03 & 1990-02-15 & 6.5 & 43.15 & $2194\pm64$ & 42.65\\
09 & PG 0953+414 & M03 & 1995-04-29 & 6.5 & 45.36 & $3155\pm44$ & 44.56\\
10 & NGC 3783 & M03 & 1993-05-23 & 4.2 & 43.14 & $3634\pm41$ & 43.02\\
11 & NGC 4151 & M03 & 1995-07-01 & 4.5 & 42.47 & $6794\pm161$ & 43.08\\
12 & PG 1226+023 & M03 & 1990-04-04 & 3.5 & 46.23 & $4023\pm174$ & 45.05\\
13 & PG 1229+204 & M03 & 1990-04-21 & 6.5 & 44.17 & $3423\pm67$ & 43.56\\
14 & PG 1307+085 & M03 & 1991-04-23 & 6.5 & 44.88 & $5382\pm187$ & 44.26\\
15 & Mrk 279 & M03 & 1989-03-26 & 5.5 & 43.85 & $5208\pm95$ & 43.30\\
16 & PG 1411+442 & M03 & 2001-06-23 & 3 & 44.49 & $2392\pm56$ & 44.08\\
17 & PG 1426+015 & M03 & 1990-02-15 & 6.5 & 44.47 & $6623\pm88$ & 44.26\\
18 & Mrk 817 & M03 & 1989-03-26 & 5.5 & 44.00 & $4937\pm120$ & 42.78\\
19 & PG 1613+658 & M03 & 1990-04-23 & 6.5 & 44.70 & $9142\pm288$ & 44.62\\
20 & PG 1617+175 & M03 & 1990-02-20 & 6.5 & 44.27 & $6445\pm327$ & 44.11\\
21 & PG 1700+518 & M03 & 1995-04-30 & 6.5 & 45.80 & $2230\pm57$ & 44.83\\
22 & 3C 390.3 & M03 & 1990-10-16 & 3.2 & 43.94 & $13665\pm264$ & 43.62\\
23 & Mrk 509 & M03 & 1996-10-12 & 3 & 44.21 & $3595\pm24$ & 43.98\\
24 & NGC 7469 & M03 & 1996-10-12 & 3 & 43.78 & $3296\pm75$ & 43.95\\
25 & Mrk 142 & B09 & 1996-11-23 & 12.5 & 43.72 & $1489\pm15$ & \nodata\\
26 & SBS 1116+583A & B09 & 1996-11-23 & 12.5 & 43.02 & $3950\pm255$ & \nodata\\
27 & Arp 151 & B09 & 1996-11-23 & 13.1 & 42.78 & $3407\pm35$ & \nodata\\
28 & Mrk 1310 & B09 & 1996-11-23 & 12.4 & 42.93 & $2731\pm51$ & \nodata\\
29 & Mrk 202 & B09 & 1996-11-23 & 12.5 & 43.04 & $1876\pm58$ & \nodata\\
30 & NGC 4253 & B09 & 1996-11-23 & 14.6 & 42.98 & $1622\pm17$ & \nodata\\
31 & NGC 4748 & B09 & 1996-11-23 & 12.5 & 43.03 & $1878\pm13$ & \nodata\\
32 & NGC 6814 & B09 & 1996-11-23 & 12.9 & 42.52 & $3202\pm13$ & \nodata\\
33 & Mrk 290 & M03 & 1990-02-16 & 6.5 & 43.44 & $5179\pm47$ & \nodata\\
34 & NGC 3227 & D10 & 2007-05-22 & 7.6 & 42.64 & $4494\pm19$ & 43.17\\
35 & NGC 3516 & D10 & 2007-05-22 & 7.6 & 43.33 & $5527\pm17$ & 43.55\\
36 & NGC 4051 & M06 & \nodata & 8 & 41.50 & $1565\pm80$ & 42.76\\
37 & NGC 5548 & B09 & 1996-11-23 & 14.7 & 43.30 & $12404\pm20$ & 43.77\\
38 & Mrk 335 & M03 & 1996-10-13 & 3 & 43.67 & $2182\pm53$ & 43.14\\
39 & Mrk 1501 & M03 & 1994-10-08 & 6.5 & 44.29 & $4959\pm37$ & \nodata\\
40 & 3C 120 & M03 & 1995-09-24 & 5$^\ast$ & 43.98 & $2419\pm29$ & 43.19\\
41 & Mrk 6 & G12 & 2010-09-15 & 7.9 & 43.63 & $4512\pm38$ & \nodata\\
42 & PG 2130+099 & M03 & 1990-09-18 & 6.5 & 44.33 & $2355\pm52$ & 42.94\\
43 & Zw 229-015 & B11a & 2010-08-15 & 5.1 & 42.93 & $4736\pm147$ & \nodata\\
44 & Mrk 50 & B11b & 2011-05-22 & 5.1 & 43.00 & $4621\pm30$ & \nodata\\
\enddata

\tablerefs{
B11a -- \citet{Barth2011a};
B11b -- \citet{Barth2011b};
B09 -- \citet{Bentz2009c};
D10 -- \citet{Denney2010};
G12 -- \citet{Grier2012};
M03 -- \citet{Marziani2003};
M06 -- \citet{Moustakas2006};
P98 -- \citet{Peterson1998}.}

\tablenotetext{$\ast$}{A typical value of 5\AA\ is assumed for 3C~120, whose FWHM$_{\rm inst}$ is not found in \citet{Marziani2003}.}

\end{deluxetable*}

\subsection{Data and Spectral Modeling}

\begin{figure*}[t]
\centering
\includegraphics[width=0.49\textwidth]{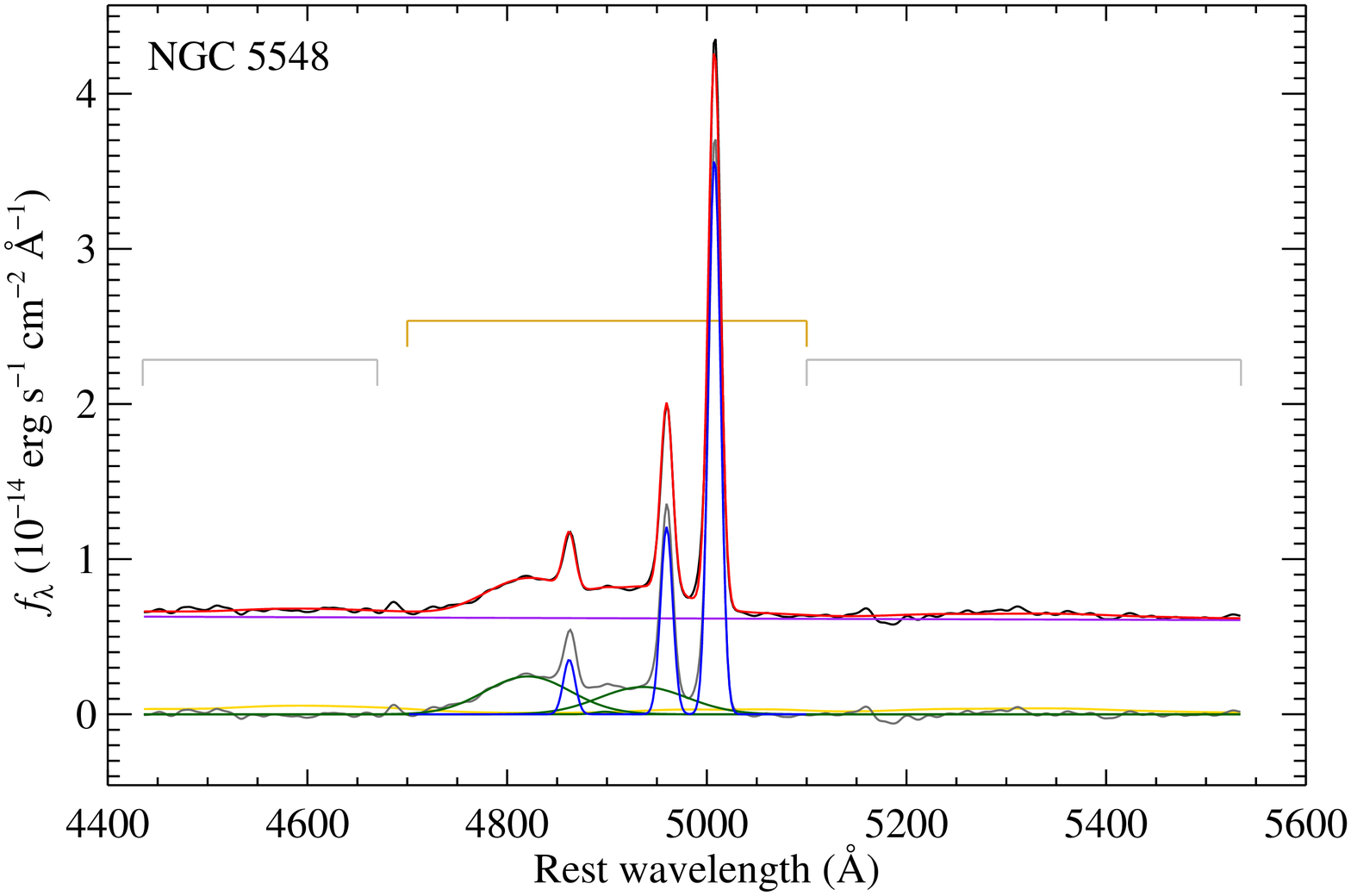}
\includegraphics[width=0.49\textwidth]{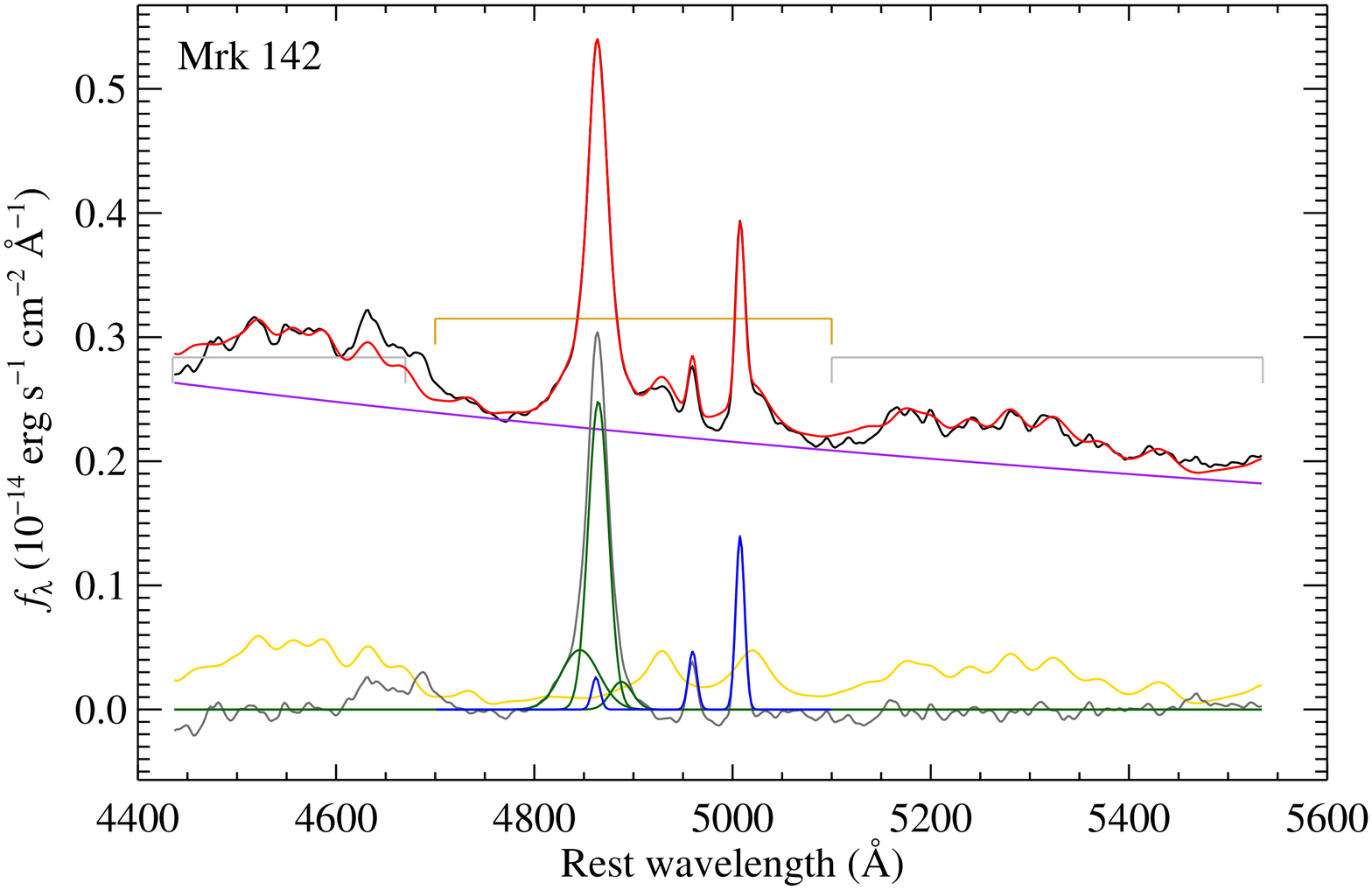}
\includegraphics[width=0.49\textwidth]{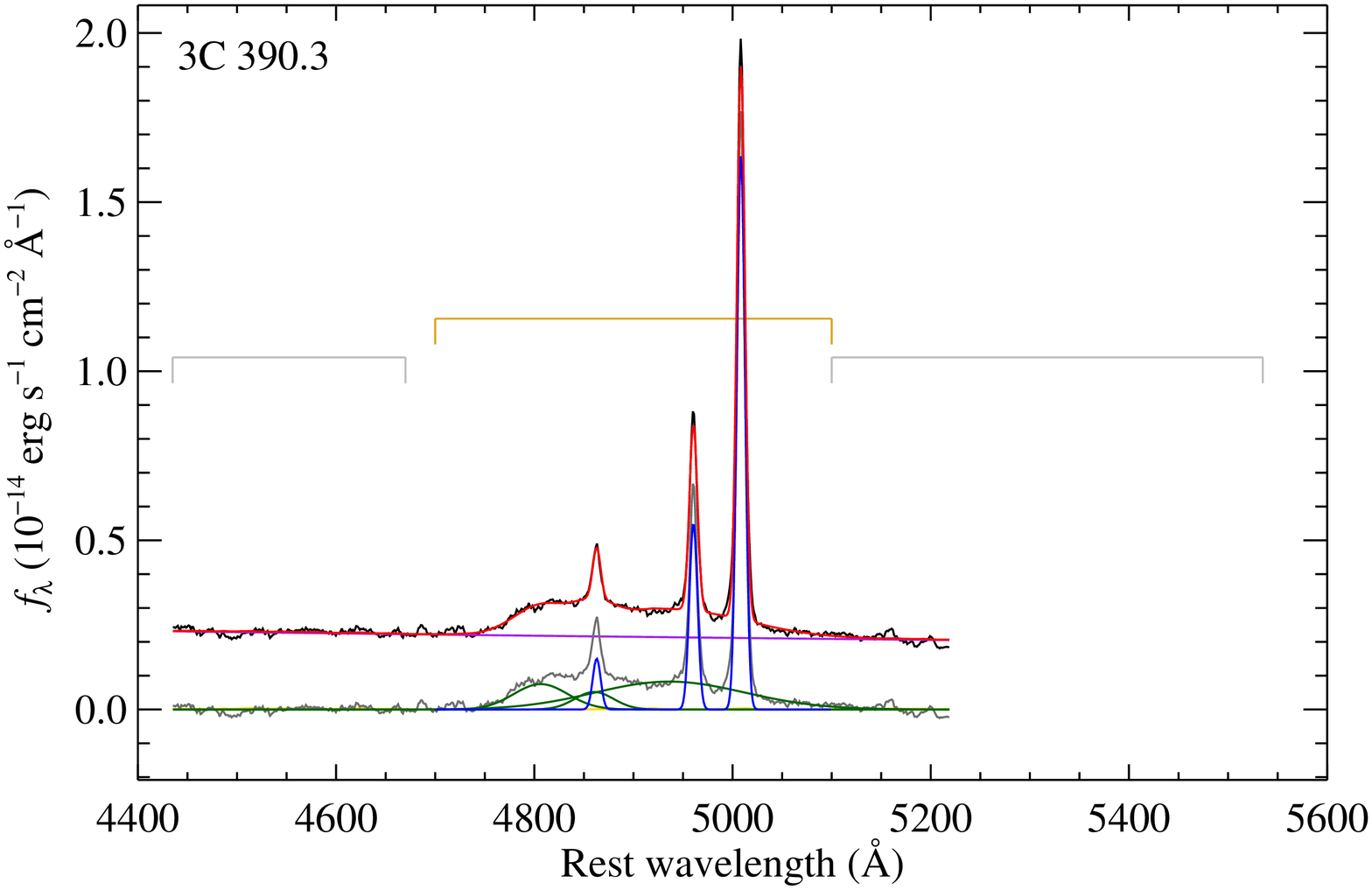}
\includegraphics[width=0.49\textwidth]{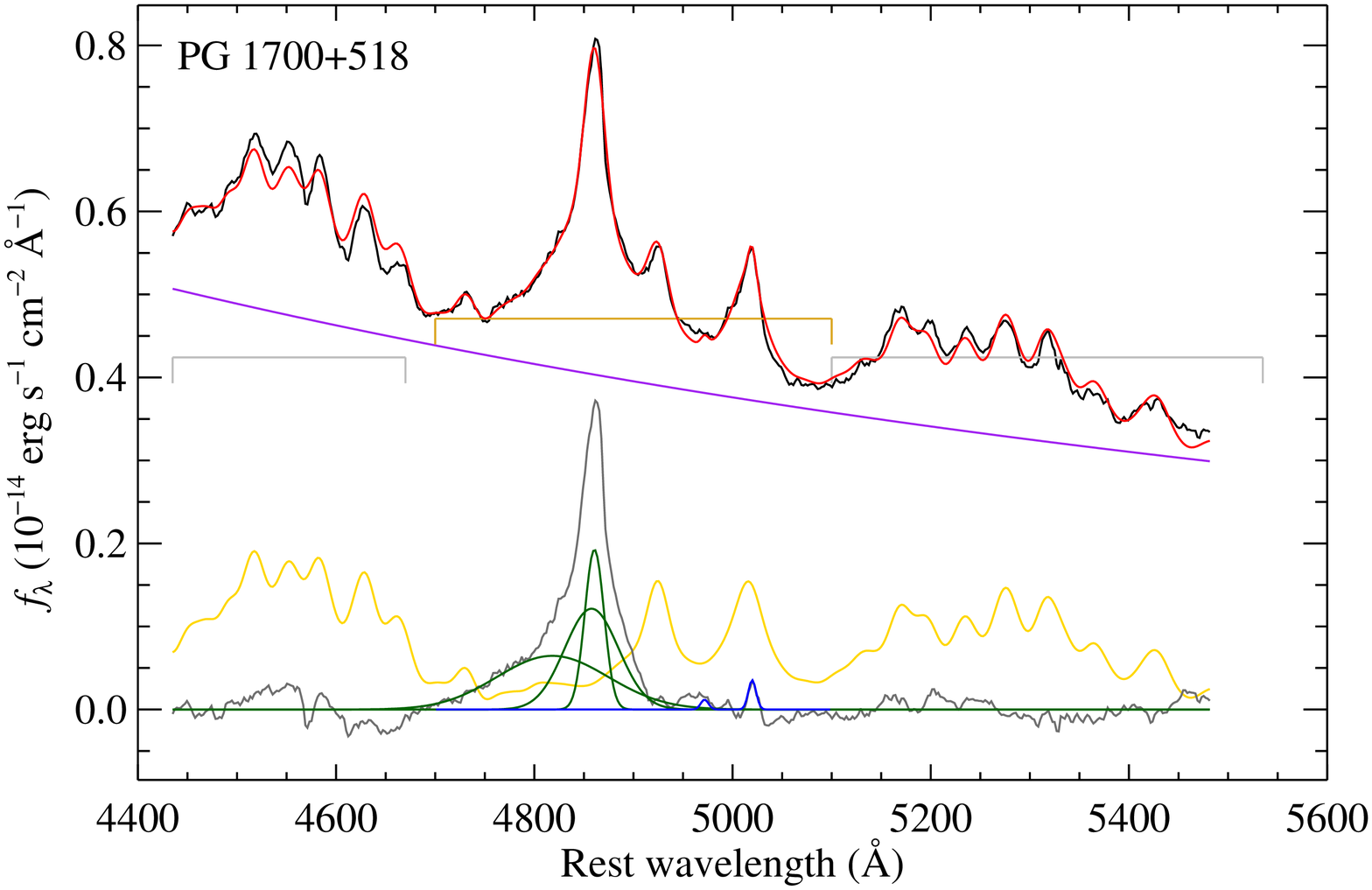}
\caption{Examples of spectral decomposition. The four panels are respectively for NGC~5548 (the best studied RM AGN), Mrk~142 (narrowest broad \hbeta\ line in our sample), 3C~390.3 (broadest \hbeta), and PG~1700+518 (strong \FeII\ and very weak narrow line emission). The black curve is the dereddened data spectrum, the red is the model spectrum, the purple is the power-law component, the yellow is the \FeII\ template, the gray is the line spectra (black $-$ purple $-$ yellow), the blue is the narrow line component, and the green is the Gaussian decomposition of the broad line component. The gray bars indicate regions where to fit the pseudo-continuum (power-law + \FeII) and the golden bar indicates the line fitting region.} 
\label{fig:specfit}
\end{figure*}

SE optical spectra covering the \hbeta\ region of the 44 RM AGN are obtained from various sources: 28 from \citet{Marziani2003}, 1 from \citet{Moustakas2006}, and the rest are provided in reduced format from the corresponding RM groups \citep{Peterson1998,Bentz2009c,Denney2010,Grier2012,Barth2011a,Barth2011b}. Only one spectrum is collected for each object, since we are interested in calibrating SE virial mass recipes. Spectral variability is then necessarily included in this calibration. We note that all spectra from the RM campaigns have been relatively corrected using the \citet{vanGroningen1992} technique and most of them are absolutely calibrated using narrow line (\OIII) flux measured from photometric nights, while those in \citet{Marziani2003} may have absolute flux calibration errors as large as 50\% \citep[see \S 2.1 in][]{Marziani2003}. The NGC 4151 spectrum from \citet{Marziani2003} seems to have a bad pixel (a sharp dip at one point) near the peak of \OIIIb. We thus interpolate at this pixel using a local fit to this narrow line with a linear continuum plus three Gaussians.

For each spectrum, the observed wavelength is converted to vacuum wavelength and shifted to the AGN rest frame, $\lambda = \lambda_{\rm obs}/(1+z)$, and the observed flux is corrected for Galactic extinction using the extinction curve of \citet{Cardelli1989} and the dust map of \citet{Schlegel1998} assuming $R_V = 3.1$. The intrinsic luminosity density is translated from the dereddened flux density, $L_\lambda = 4 \pi d_L^2 f(1+z)$. 

Following \citet{Shen_etal_2008}, we decompose each rest-frame spectrum into various emission components. The fitting procedure is briefly described here and more details can be found in earlier work \citep[e.g.,][]{Shen_etal_2008,Shen2011}. A pseudo-continuum, which is a combination of a power-law component and an \FeII\ template constructed from the narrow line Seyfert 1 galaxy I Zw 1 \citep{Boroson1992}, is first fitted in the rest-frame 4435--4670\AA\ and 5100--5535\AA. The continuum-subtracted line spectrum is then fitted with 6 Gaussian components, 3 for narrow lines (\hbetawave\ and \OIIIab) and 3 for the broad \hbeta, respectively, in the rest-frame wavelength region of 4700\AA--5100\AA. All narrow lines are constrained to have the same shift (relative to systemic) and line width. The broad \HeIIwave\ region is avoided if the emission line appears to be evident in the residual. Four examples of spectral decomposition are shown in Figure~\ref{fig:specfit} for illustration; others have similar data quality and goodness of fit. The monochromatic continuum luminosity, $L_{5100} = \lambda L_\lambda$ at 5100\AA, is quoted as that of the power-law component.  The FWHM of the broad \hbeta\ is computed from the model spectra of broad lines, and corrected for instrumental broadening. The errors of $L_{5100}$ and FWHM are estimated from 100 mock spectra generated by adding random Gaussian noise at each spectral pixel based on quoted flux errors. The spectral measurements are listed in Table~\ref{tab:spec}. We do not list the measurement errors for $\log L_{5100}$ as they are very small, all less than 0.01 dex with a median of 0.0025 dex. 

\section{Recalibration of SE Virial Mass Estimators}
\label{sec:recal}

\subsection{Bases of the SE Estimator}

Two substitutions are made from the RM to SE mass estimator. The first is to use $L$ for $R$ based on the radius-luminosity ($R-L$) relation. The second is to substitute the rms line dispersion $\sigma_{\rm rms}$ with the line width measured from SE spectra. As the measurement of line dispersion relies more on the details of  spectral decomposition (e.g., the choice of models and \FeII\ templates) and may lead to less repeatable results, we will use FWHM for the SE line width in this work. We first calibrate the coefficients on $L$ and FWHM in the SE mass estimator (Equation~\ref{eqn:SE}).

The bivariate correlated errors and intrinsic scatter (BCES) algorithm \citep{Akritas1996} is used for linear regression when measurement errors on both $X$ and $Y$ need be taken into account. In the case where the measurement errors on $X$ are much smaller than on $Y$, we use the weighted least squares (WLS) method described in \citet{Akritas1996}. Both regression techniques incorporate intrinsic and measurement variances. To differentiate from the conventional WLS, we call it intrinsic and measurement variances weighted least squares (IMVWLS). The calculation of the total and intrinsic variances after BCES regression, and an extension of IMVWLS from the bivariate to multivariate case are described in the Appendix. For asymmetric error bars, the one on the side closer to the regression line is chosen; a few iterations may be needed until convergence. 

\begin{figure}[t]
\centering
\includegraphics[width=\columnwidth]{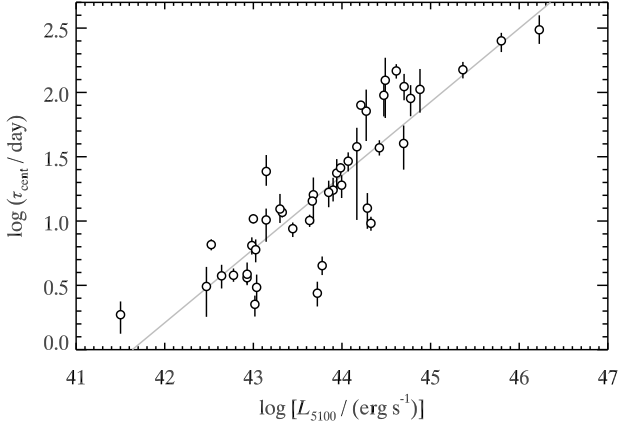}
\caption{Time delay ($\tau_{\rm cent}$) versus the continuum luminosity $L_{5100}$, i.e.\ the radius-luminosity relation $R \propto L^\alpha$, with the regression line (Equation~\ref{eq:rl}). The errors on $\log L_{5100}$ are smaller than the symbol.
\label{fig:rl}}
\end{figure}

The best-fit $R-L$ relation ($R \propto L^\alpha$) of our sample is shown in Figure~\ref{fig:rl}, which is 
\begin{equation}
\label{eq:rl}
\log\left(\frac{\tau_{\rm cent}}{{\rm day}}\right) = 
  -23.82 \pm 1.97 + (0.572 \pm 0.045) 
  \log \left( \frac{L_{5100}}{\rm erg\; s^{-1}} \right)
\end{equation}
found with IMVWLS because the measurement errors on $L_{5100}$ is much smaller than those on $\tau_{\rm cent}$. The slope $\alpha$ is shallower than the previous result obtained from a smaller sample \citep[$0.67 \pm 0.05$ found by][]{Kaspi2005}, steeper than but consistent within 1-$\sigma$ with that corrected for host starlight \citep[$0.533 \pm 0.034$ found by][]{Bentz2013}. We estimated a total scatter $\sigma_{\rm tot} = 0.29$ dex and an intrinsic scatter $\sigma_{\rm int} = 0.25$ of the best-fit $R-L$ relation shown in Figure~\ref{fig:rl}.

\begin{figure}[h]
\centering
\includegraphics[width=\columnwidth]{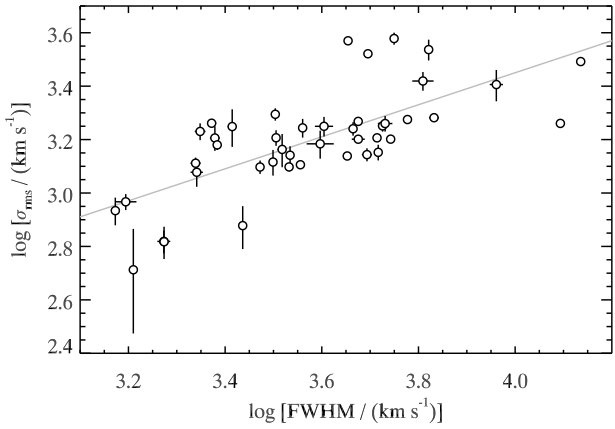}
\caption{Correlations between $\sigma_{\rm rms}$ and SE FWHM. The solid line is the regression line, see Equation~(\ref{eq:fwhm}).
\label{fig:linewidth}}
\end{figure}

Figure~\ref{fig:linewidth} shows the comparison between $\sigma_{\rm rms}$ and SE FWHM for our RM sample. The linear regression with BCES$(\sigma_{\rm rms} | {\rm FWHM})$ gives 
\begin{equation}
\label{eq:fwhm}
\log \left(\frac{\sigma_{\rm rms}}{\rm km\;s^{-1}}\right)
  = 1.052 \pm 0.360 + (0.600 \pm 0.100) \log \left(\frac{\rm FWHM}{\rm km\;s^{-1}}\right)
\end{equation}
with a total scatter of 0.14 dex and an intrinsic scatter of 0.13 dex. Our best-fit slope is consistent with the slope of $0.59 \pm 0.06$ \citep{Collin2006} or $0.54 \pm 0.08$ \citep{Wang2009}, where FWHM is measured from the mean spectra. A correlation is clearly seen, although there is a significant scatter around the mean relation, and the slope of this relation is not unity. This justifies the usage of FWHM in SE mass estimators, but argues for a different dependence on FWHM for SE mass estimators.  

Here we note that the calibration of these two relations does not take the AGN variability into account. In our sample, there are only 14 out of 44 SE spectra were taken contemporaneously during the course of the RM measurement. We argue that this may introduce an additional but limited scatter into the relations. Taking NGC 5548 as an example, which is one of the most variable AGN in the RM sample, the rms variability is found to be 0.1~dex for $L_{5100}$ and 0.07~dex for FWHM based on 241 observations made from December 1988 to September 1993\footnote{The revised selected optical spectra for NGC 5548 in the AGNwatch program available at http://www.astronomy.ohio-state.edu/$\sim$agnwatch/}. For most other objects, the variability is smaller. Our results (slopes of the two relations) are well consistent with those computed using contemporaneous data within 1 $\sigma$, \citep[cf.][]{Bentz2013,Collin2006,Wang2009}, further suggesting that including non-contemporaneous RM and SE data does not degrade the best-fit relations much; see \S\ref{sec:hostcorr} for the comparison with the host light corrected $R-L$ relation.

\subsection{Recalibration of SE virial mass estimators}

\begin{deluxetable*}{ccrrcrrrr}
\tablecolumns{9}
\tablewidth{0pc}
\tablecaption{The slope and scatter after linear regression, and the zeropoint between $\log M_{\rm RM}$ and $\log \mu$ with two different recipes. \label{tab:reg}}

\tablehead{
\colhead{} & \colhead{} &
\multicolumn{2}{c}{BCES(bisector)} & \colhead{} & 
\multicolumn{2}{c}{IMVWLS} & \colhead{} \\
\cline{3-4} \cline{6-7} \\
\colhead{$\alpha$} & \colhead{$\beta$} &
\colhead{$b$} & \colhead{$\sigma_{\rm tot}$/$\sigma_{\rm int}$} & \colhead{} &
\colhead{$b$} & \colhead{$\sigma_{\rm tot}$/$\sigma_{\rm int}$} &
\colhead{zeropoint} & \colhead{$\sigma_{\rm int}$}}
\startdata
\multicolumn{9}{c}{without host light correction}\\
\noalign{\smallskip}\hline\noalign{\smallskip}
0.500 & 2.000 & $1.18\pm0.10$ & 0.46/0.42 & & $1.01\pm0.10$ & 0.45/0.41 & $0.613\pm0.013$ & 0.41\\
0.572 & 1.200 & $1.30\pm0.11$ & 0.39/0.35 & & $1.16\pm0.09$ & 0.39/0.35 & $3.495\pm0.012$ & 0.36\\
\cutinhead{with host light correction}
0.500 & 2.000 & $1.08\pm0.11$ & 0.45/0.40 & & $0.93\pm0.09$ & 0.43/0.40 & $0.733\pm0.021$ & 0.39\\
0.504 & 1.200 & $1.28\pm0.13$ & 0.41/0.36 & & $1.12\pm0.09$ & 0.40/0.36 & $3.602\pm0.021$ & 0.35\\
\enddata

\end{deluxetable*}

\begin{figure}[t]
\centering
\includegraphics[width=0.8\columnwidth]{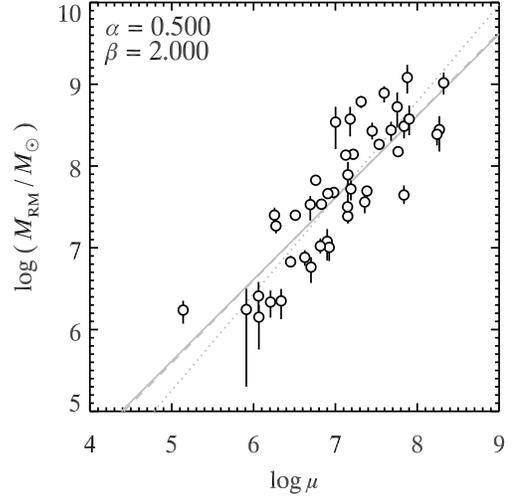}\\
\includegraphics[width=0.8\columnwidth]{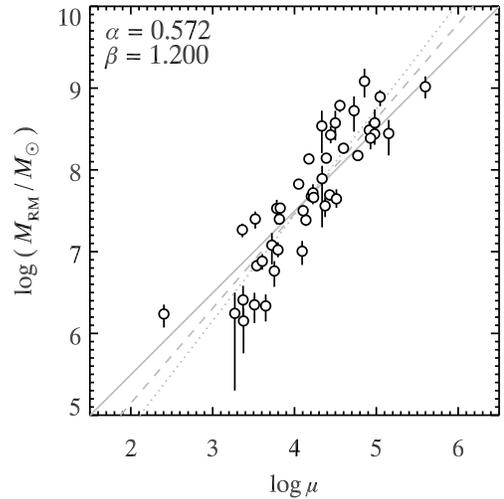}
\caption{RM masses versus SE virial products defined in Equation~\ref{eq:sevp}.  The solid line is for $\log\left(M_{\rm SE}/M_\sun\right) = \log \mu + {\rm zeropoint}$, the dotted line is a result of BCES(bisector), and the dashed line is from IMVWLS.
\label{fig:reg}}
\end{figure}

Now we are ready to calibrate the SE mass estimator in Equation~(\ref{eqn:SE}). The SE virial product is defined as
\begin{equation}
\label{eq:sevp}
\mu = \left(\frac{L_{5100}}{\rm 10^{44} \; erg\; s^{-1}}\right)^\alpha
  \left(\frac{{\rm FWHM}}{\rm km\; s^{-1}}\right)^\beta \; ,
\end{equation}
where $\alpha$ could be either 0.5 (from simple photoionization prediction) or 0.572 (from our regression fit, Equation~\ref{eq:rl}), and $\beta$ could be either 2 (assuming FWHM is a virial velocity indicator) or $2 \times 0.600$ (from our regression fit, Equation~\ref{eq:fwhm}). Following \citet{Vestergaard2006}, the SE mass can be written as
\begin{equation}
\label{eq:semass_zp}
\log\left(\frac{M_{\rm SE}}{M_\sun}\right) = \log\left(\mu\right) + {\rm zeropoint},
\end{equation}
where the zeropoint is determined as a weighted mean of $\log(M_{\rm RM}) - \log(\mu)$.  Following IMVWLS, here both intrinsic and measurement errors are included as weights.

To test whether or not the SE virial product can be used as an unbiased estimator of the RM mass, we first check if $\mu$ is linearly correlated with $M_{\rm RM}$ using BCES(bisector). Because the measurement errors on $M_{\rm RM}$ is in general much larger than on $\mu$,  regression with IMVWLS, which belongs to the type $Y|X$, is also employed.  Two combinations of $\alpha$ and $\beta$ as mentioned above are tested, respectively, with results listed in Table~\ref{tab:reg} and shown in Figure~\ref{fig:reg}. Both recipes result in a linear relation between the SE virial product and the RM mass, i.e.\ the slope $b$ is found to be consistent with unity within 1 or 2$\sigma$. It is seen that the best-fit slopes $\alpha = 0.572$ and $\beta = 1.200$ result in a smaller intrinsic scatter ($\sigma_{\rm int} = 0.36$) than using the conventional values: $\alpha = 0.5$ and $\beta = 2$ ($\sigma_{\rm int} = 0.41$).

\subsection{Host galaxy contribution to the AGN luminosity}
\label{sec:hostcorr}

The observed AGN flux is more or less contaminated by starlight from the host galaxy. The fractional contamination is more severe for nearby low-luminosity objects; such effect will change the power-law slope of the $R-L$ relation, which is the cornerstone of SE techniques. A flatter $R-L$ relation is obtained if the host light contribution is appropriately removed. \citet{Bentz2013} found a slope of $0.533 \pm 0.034$ after host light correction, much smaller than that without correction \citep[e.g., $0.67 \pm 0.05$ found by][]{Kaspi2005}. 

\begin{figure}[bth]
\centering
\includegraphics[width=0.8\columnwidth]{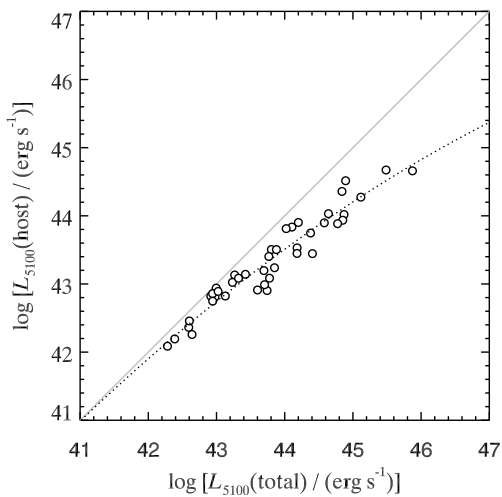}
\caption{Host versus total luminosity ($\lambda L_\lambda$ at 5100\AA) in the spectra of RM AGN from the sample of \citet{Bentz2013}. The error bars are smaller than the symbols. The solid line is the 1:1 diagonal and the dotted line is a binomial relation that fits the data. 
\label{fig:hostcorr}}
\end{figure}

We plot the host luminosity versus the total luminosity in the aperture using RM data adopted from \citet{Bentz2013}, see Figure~\ref{fig:hostcorr}. To avoid giving complicated weights, we choose only one observation for each object if there are many, the one with the smallest slit width. This is because RM observations usually favor larger apertures than usual. It is obvious that the host contribution is scaled with the AGN luminosity, probably due to the correlation between the black hole mass and the bulge mass. A linear function is able to adequately characterize the relation. However, it intercepts the diagonal at $\sim$$10^{42}$~\ergs, making the relation invalid for AGN fainter than this luminosity. We thus fit the the data using a binomial with one point fixed at $(40, 40)$ in Figure~\ref{fig:hostcorr}, assuming that no AGN is fainter than $\sim 10^{40}$~\ergs. This gives an empirical estimate of the host luminosity in AGN spectra at 5100\AA,
\begin{equation}
\label{eq:hostcorr}
\begin{split}
& \log \left[ \frac{L_{5100} ({\rm host})}{\rm erg\; s^{-1}} \right]
 = 1.024 \times \log \left[ \frac{L_{5100} ({\rm total})}{\rm 10^{40}\; erg\; s^{-1}} \right] \\
&\quad -0.0367 \times \left\{ \log \left[ \frac{L_{5100} ({\rm total})}{\rm 10^{40}\; erg\; s^{-1}} \right] \right\}^2
  + 40 \; .
\end{split}  
\end{equation}
The 2nd order item is needed at a significance level of 0.9998 with F-test, while a 3rd order item is rejected at a significance level of 0.5. The total variance after regression is 0.0350 (=0.187~dex), quoted as the uncertainty for prediction. The uncertainty is mainly due to AGN variability and host diversity, and moderately dependent on the aperture size and seeing, because a small aperture will reduce both the host and AGN fluxes in varying degrees. Thus, such relation can be used for estimation of the fractional host contamination given the total luminosity with a precision of 0.19~dex. We emphasize that this relation is {\it only} valid in the luminosity range of $10^{42}-10^{46}$~\ergs, and for hosts similar to those in the current RM sample; Extrapolating to low luminosities should be cautious, while extrapolation to high luminosities is safe, as the host contamination is negligible anyway. 

\begin{figure}[tb]
\centering
\includegraphics[width=\columnwidth]{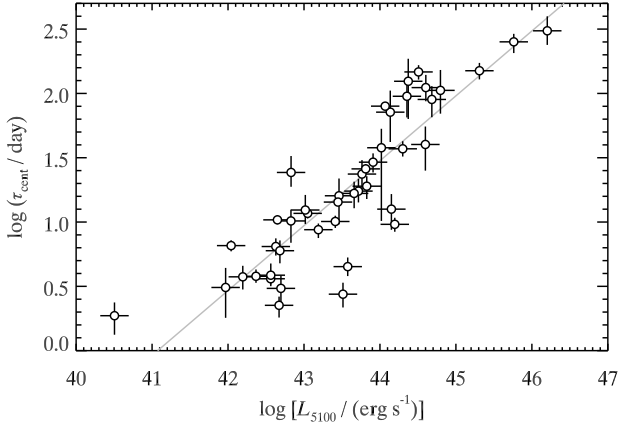}
\caption{Same as Figure~\ref{fig:rl} but using host light corrected $L_{5100}$. The regression line is expressed in Equation~\ref{eq:rlhost}. The errors on $\log L_{5100}$ are mainly due to uncertainties in host light correction (Equation~\ref{eq:hostcorr}).
\label{fig:rlhost}}
\end{figure}
\begin{figure}[ht]
\centering
\includegraphics[width=0.8\columnwidth]{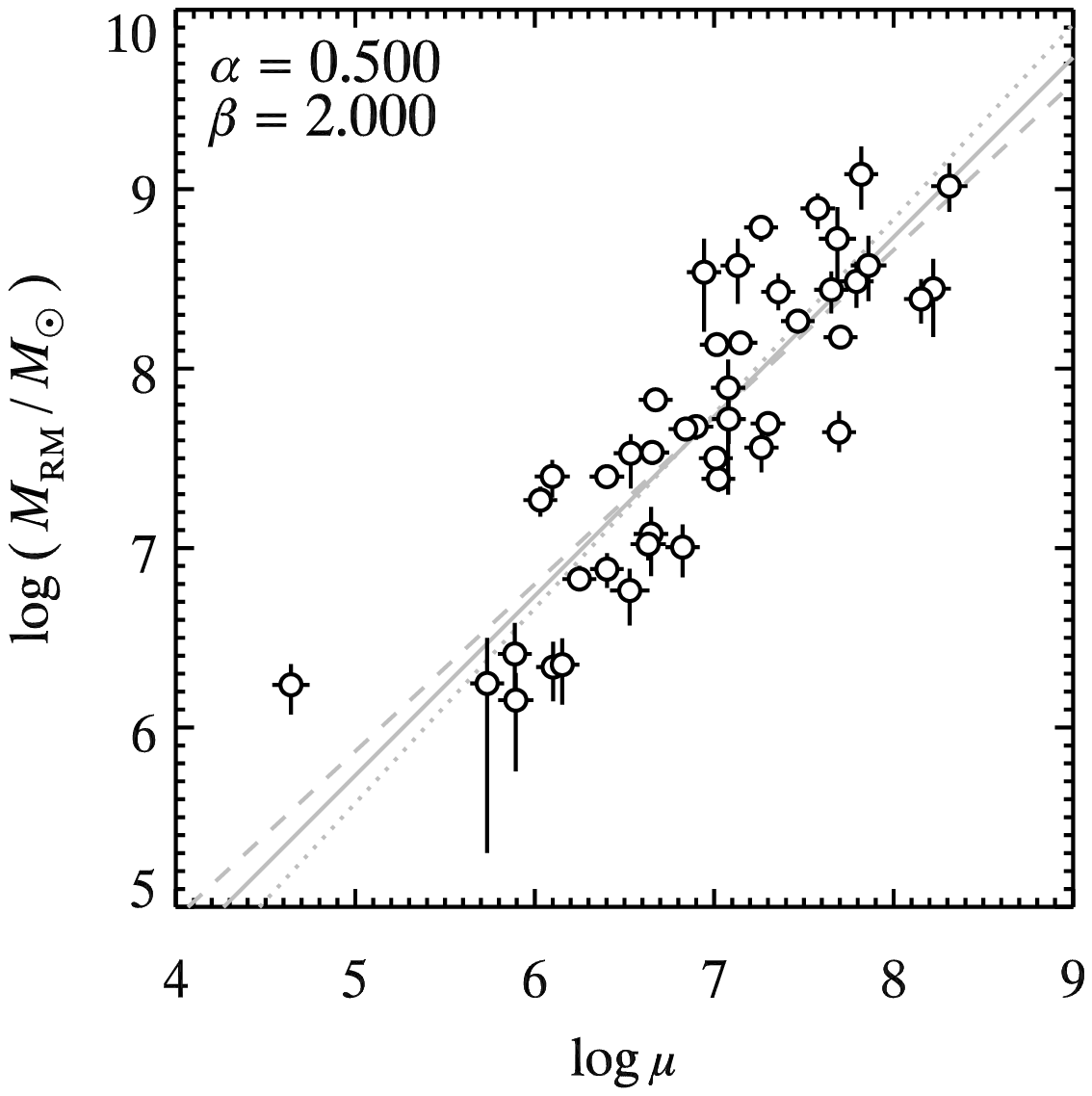}\\
\includegraphics[width=0.8\columnwidth]{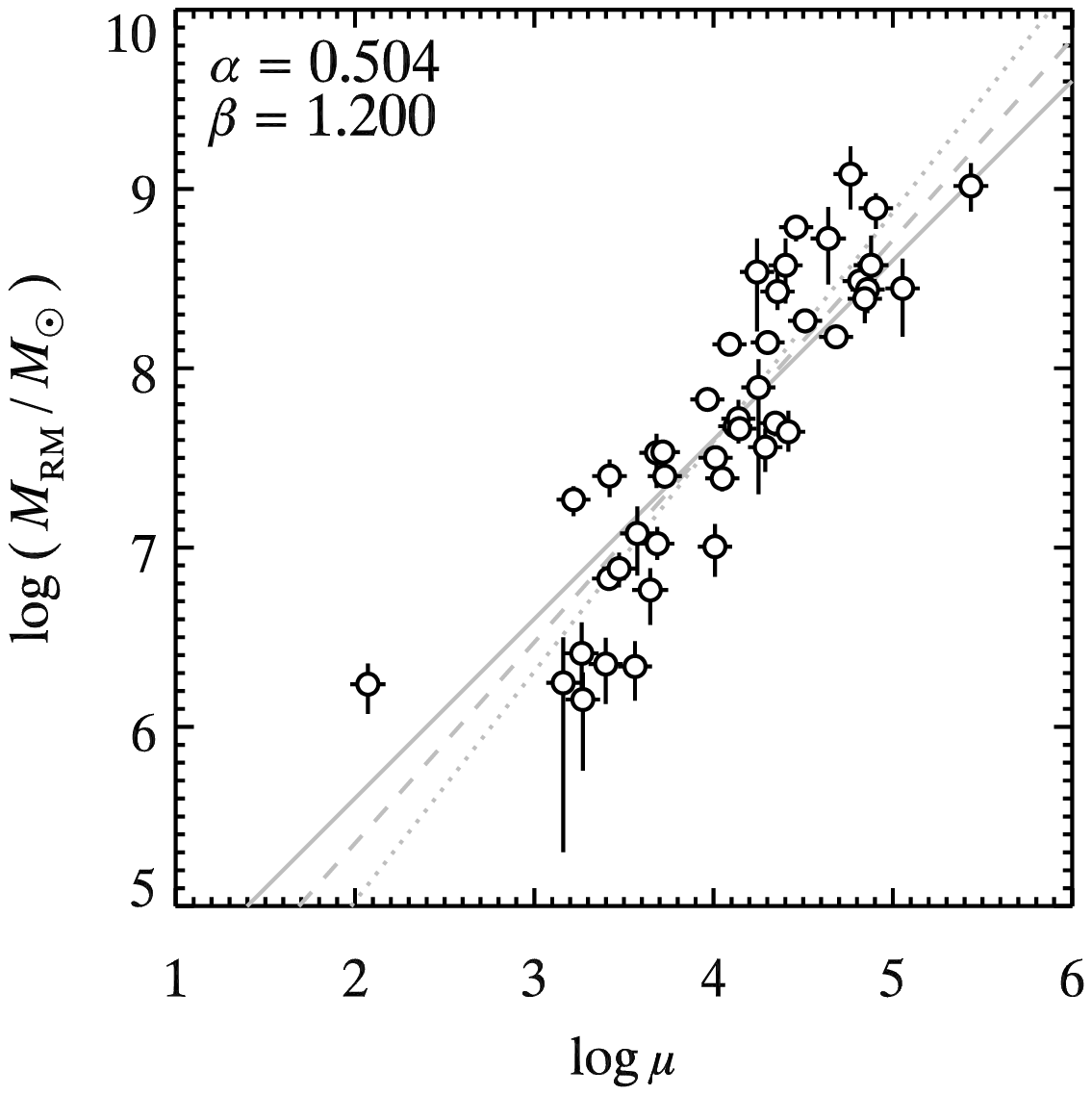}
\caption{Same as Figure~\ref{fig:reg} but using host light corrected $L_{5100}$ and corresponding $\alpha$ and $\beta$.
\label{fig:regcorr}}
\end{figure}
\begin{figure}[ht]
\centering
\includegraphics[width=\columnwidth]{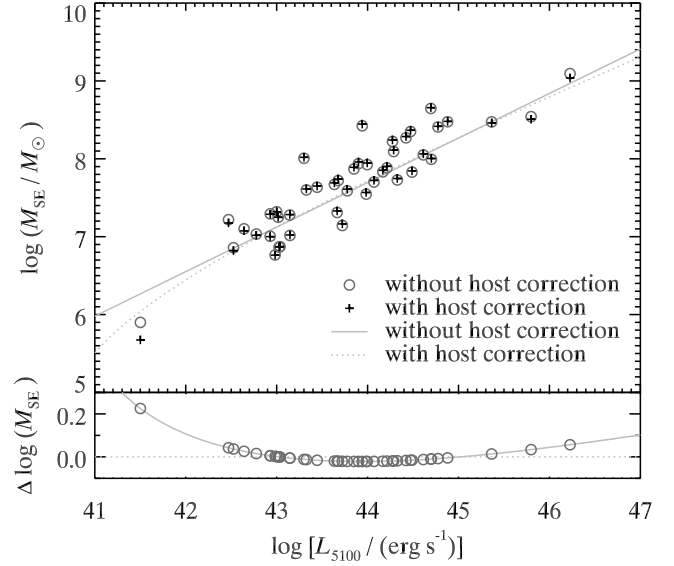}
\caption{Comparison of SE masses derived using the recipe with direct $L_{5100}$ and the one using host light corrected $L_{5100}$. The points are objects in our sample and the lines are SE masses assuming ${\rm FWHM} = 10^{3.5}$~km~s$^{-1}$. The lower panel shows the difference of the two mass recipes in unit of dex.
\label{fig:cmp_hostcorr}}
\end{figure}

Applying Equation~\ref{eq:hostcorr} to the SE luminosities for all objects in our sample, the $R-L$ relation is recalibrated to be
\begin{equation}
\label{eq:rlhost}
\log\left(\frac{\tau_{\rm cent}}{{\rm day}}\right) = 
  -20.72 \pm 1.91 + (0.504 \pm 0.044) 
  \log \left( \frac{L_{5100}}{\rm erg\; s^{-1}} \right),
\end{equation}
using BCES$(\tau_{\rm cent}|L_{5100})$ with a total scatter of 0.30~dex and an intrinsic scatter of 0.24~dex, see Figure~\ref{fig:rlhost}. With this empirical host light correction, the best-fit slope on the $R-L$ relation is well consistent with that measured by \citet{Bentz2013}. Using the new $R-L$ relation with host contamination corrected $L_{5100}$, the virial SE estimators are tested and shown in Table~\ref{tab:reg} and Figure~\ref{fig:regcorr}.

The estimated SE masses derived from the two recipes, one using direct (total) $L_{5100}$ and the other using host light corrected $L_{5100}$, are compared in Figure~\ref{fig:cmp_hostcorr}. For objects in our sample, NGC 4051 is the only object displaying a large discrepancy, 0.23~dex, as our direct $L_{5100}$ is an order of magnitude lower than that quoted in \citet{Bentz2013}, $10^{41.5}$ versus $10^{42.4}$~\ergs, probably due to AGN variability and a small aperture used in \citet{Moustakas2006}. The SE masses for other objects are quite similar, with a maximum difference of 0.056~dex and a median of 0.017~dex. The lines show SE masses from the two recipes against luminosity assuming a line width ${\rm FWHM} = 10^{3.5}$~km~s$^{-1}$. As the two recipes use the same $\beta$ value, their discrepancy is independent of the broad line width. The difference of the two recipes becomes large at low luminosities, about 0.1~dex at $10^{42}$~\ergs\ and 0.5~dex at $10^{41}$~\ergs. At the high luminosity end, both recipes produce consistent results, with a discrepancy of 0.1~dex at $10^{47}$~\ergs. Between $10^{42}$ and $10^{47}$~\ergs, the two recipes produces almost identical results. This is the luminosity range containing most of the objects in this study. Thus, we choose to adopt the recipe using direct $L_{5100}$ in the following of this work.

\section{A New Technique for AGN Mass Estimate}
\label{sec:opt}

Spectral decomposition and line width measurement are more or less dependent on the choice of models and details of the fitting procedure. The results could be problematic under extreme conditions, for instance, at low $S/N$. In these cases, the flux or luminosity may be the only quantity that can be robustly measured among all spectral properties. We thus aim to find a scaling relation between luminosities and the RM mass using optimization techniques, inspired by a work in chemometrics \citep{Wu2009}.

\subsection{Optimization Setup}

We define $k$ rectangle filters and let $\lambda_{i,1}$ and $\lambda_{i,2}$ be the left and right bounds, respectively, of the $i$th filter. Provided the spectrum $L_{\lambda}$ of an object, the luminosity under the $i$th filter is given by
\begin{equation}
L_{i} = \int_{\lambda_{i,1}}^{\lambda_{i,2}}{L_{\lambda}} d\lambda
  \quad (i = 1, ..., k) \; .
\end{equation}
Our goal is to find a set of filters and coefficients $b_i$ to build a tight and robust relation between the filtered luminosities and black hole masses using optimization techniques,
\begin{equation}
\log(M_{\rm BH}) = b_0 + \sum\limits_{i=1}^{k} b_i \log(L_i) \; .
\end{equation}

In each iteration of the optimization, the filters are searched in a defined parameter space, and the coefficients $b_i$ are computed using IMVWLS (see Appendix \ref{sec:appb}) given the RM masses ($M_{\rm RM}$) and filtered luminosities ($L_i$). The objective of the optimization is to find a set of filters that results in a minimum intrinsic scatter ($\sigma_{\rm int}$) after IMVWLS regression between $\log M_{\rm RM}$ and $\log L_i$.

We divide the whole sample into two subsets: the calibration and prediction sets with $n_c$ and $n_p$ objects, respectively. The calibration set is used to search for the optimal filers and compute the coefficients $b_i$. The prediction set is not used in finding optimal filters and coefficients, but provides an independent test of the results. 

Following \citet{Wu2009}, we adopt the particle swarm optimization (PSO) technique that has been proven powerful and robust in finding the global minimum. Other techniques such as the genetic algorithm should work as well. The PSOt toolbox\footnote{Available at \url{http://www.mathworks.com/matlabcentral/fileexchange/7506}} implemented in MATLAB is used in practice thanks to its friendly interface and fast computation. The PSO configuration includes: a particle population size of 60, maximum velocity equal to 1/2 of the parameter range, an inertia constant of 0.6 and both acceleration coefficients of 1.7 \citep[type 1 tuning in][]{Trelea2003}. The search stops if the global minimum does not change by more than 1E-200 in 100 iterations. How the PSO works and the meaning of these parameters can be found in \citet{Wu2009} and \citet{Trelea2003} and references therein. The above details are sufficient for the reader to repeat our work and it is beyond the scope of this paper to explain more details on this PSO technique.  

All optical spectra have a common coverage in the wavelength range of 4700--5200\AA\ at rest frame. We thus choose this region for optimization. For the purpose of fast computation in MATLAB, we re-sample all optical spectra from 4700\AA\ to 5200\AA\ with a step of 0.5\AA\ to create a uniform data set. Each filter has two independent parameters, its central wavelength and width. The central wavelength is searched in the overall range of the spectrum, i.e.\ 4700--5200\AA, and the width is confined in the range of 5--500\AA. 

\begin{figure}[t]
\centering
\includegraphics[width=\columnwidth]{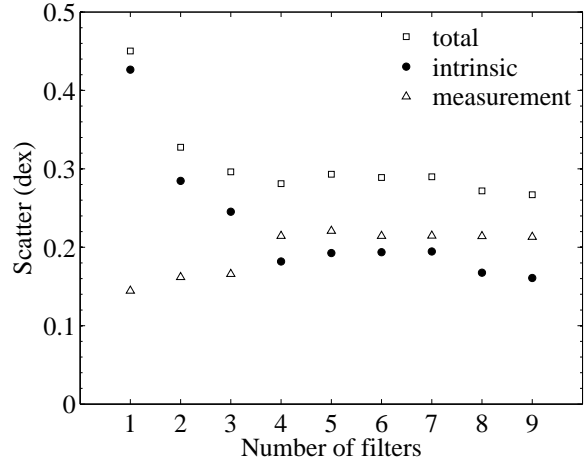}
\caption{Total and intrinsic scatters and the average measurement errors after IMVWLS regression versus the number of filters in the case with all objects in the calibration set ($n_c=44$). The mean measurement error ($=\sqrt{\bracket{ \sigma_y^2}}$) varies because the error bars are asymmetric.
\label{fig:nfilter}}
\end{figure}

\subsection{Results}

In the first trial, we put all objects in the calibration set ($n_c = 44$ and $n_p = 0$). As expected, the scatter after regression decreases with increasing number of filters, see Figure~\ref{fig:nfilter}. However, too many filters may over-fit the data, i.e.\ fitting the noise. If this happens, the mass estimate will result in a large uncertainty in prediction, amplified by noise fluctuation.  The partial F-test is employed to test the significance of adding the $(k+1)$th filter with respect to the model with $k$ filters. The chance probability is $1.3 \times 10^{-6}$ for the 2nd filter, and larger than $10^{-2}$ for any addition. Therefore, any filters in addition to the 2nd should not be added at a significance level of 0.01. We thus conclude that the optimal number of filters is 2.

\begin{figure}[t]
\centering
\includegraphics[width=\columnwidth]{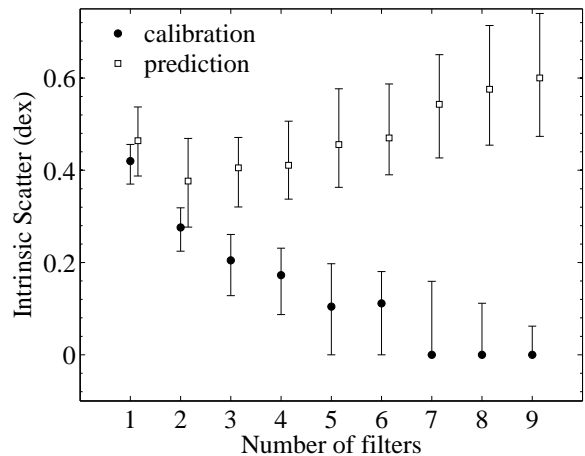}
\caption{Intrinsic scatter for calibration and prediction versus number of filters. Given a number of filters, the members of each set ($n_c = 29$ and $n_p = 15$) are randomly selected and repeated for 100 times to calculate the median value and the 68\% error range, shown as the points and bars, respectively. The intrinsic scatter is set to zero if the total variance after regression is smaller than the measurement variance. The prediction data are shifted horizontally for clarity. 
\label{fig:rand}}
\end{figure}
\begin{deluxetable}{cccr}
\tablecolumns{4}
\tablewidth{0pc}
\tablecaption{Positions of the two optimal filters and the coefficients from IMVWLS.
\label{tab:filt}}
\tablehead{
\colhead{$i$} & \colhead{$\lambda_{i,1}$} & \colhead{$\lambda_{i,2}$} & \colhead{$b_i$}
}
\startdata
0 & \nodata & \nodata &          $-21.55 \pm 2.12$ \\
1 &	    4761.8 &	    4844.6 &	      $9.89 \pm 1.43$ \\
2 &	    4727.3 &	    4751.2 &	     $-9.33 \pm 1.45$ \\
\enddata
\end{deluxetable}

To further investigate the choice of the best model, we split the whole sample into a calibration set and a prediction set, as described above, to provide a true estimation of its prediction capability. After a few attempts, we decide to choose $n_c = 29$ and $n_p = 15$ for both sets to have sufficient members. Varying the number by a few will not affect the results.  The selection of individual members for the calibration set is critical. The calibration members should well sample the diversity of AGN, i.e., evenly populated in the space of accretion rate, ionization state, host contamination, etc., to cover all aspects that may affect the shape of the optical spectrum of AGN. In other words, one always wish to make interpolations rather than extrapolations for prediction. As it is hard to make an optimal selection, we decide to randomly choose the calibration members and repeat 100 times to smooth out the bias caused by the selection effect. We note that the median value of the intrinsic scatter from the 100 times of randomization is a typical but not an optimal result. As shown in Figure~\ref{fig:rand}, the intrinsic scatter $\sigma_{\rm int}$ for prediction increases gradually when there are 2 or more filters. As discussed already, one starts to over-fit the data with 3 or more filters, and the optimal number of filters is 2, consistent with the conclusion from the partial F-test.

\begin{figure}[t]
\centering
\includegraphics[width=0.8\columnwidth]{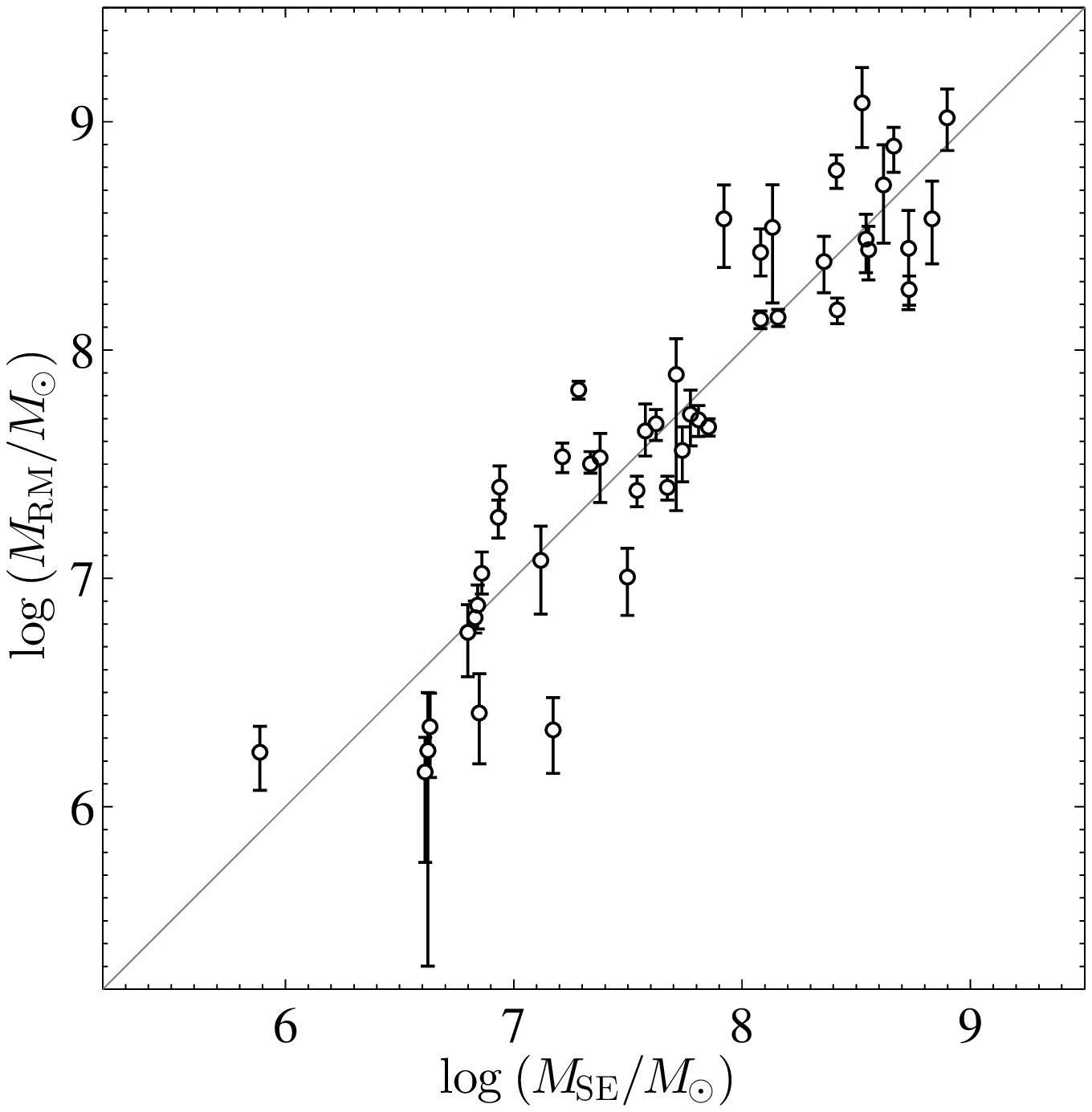}\\
\includegraphics[width=0.9\columnwidth]{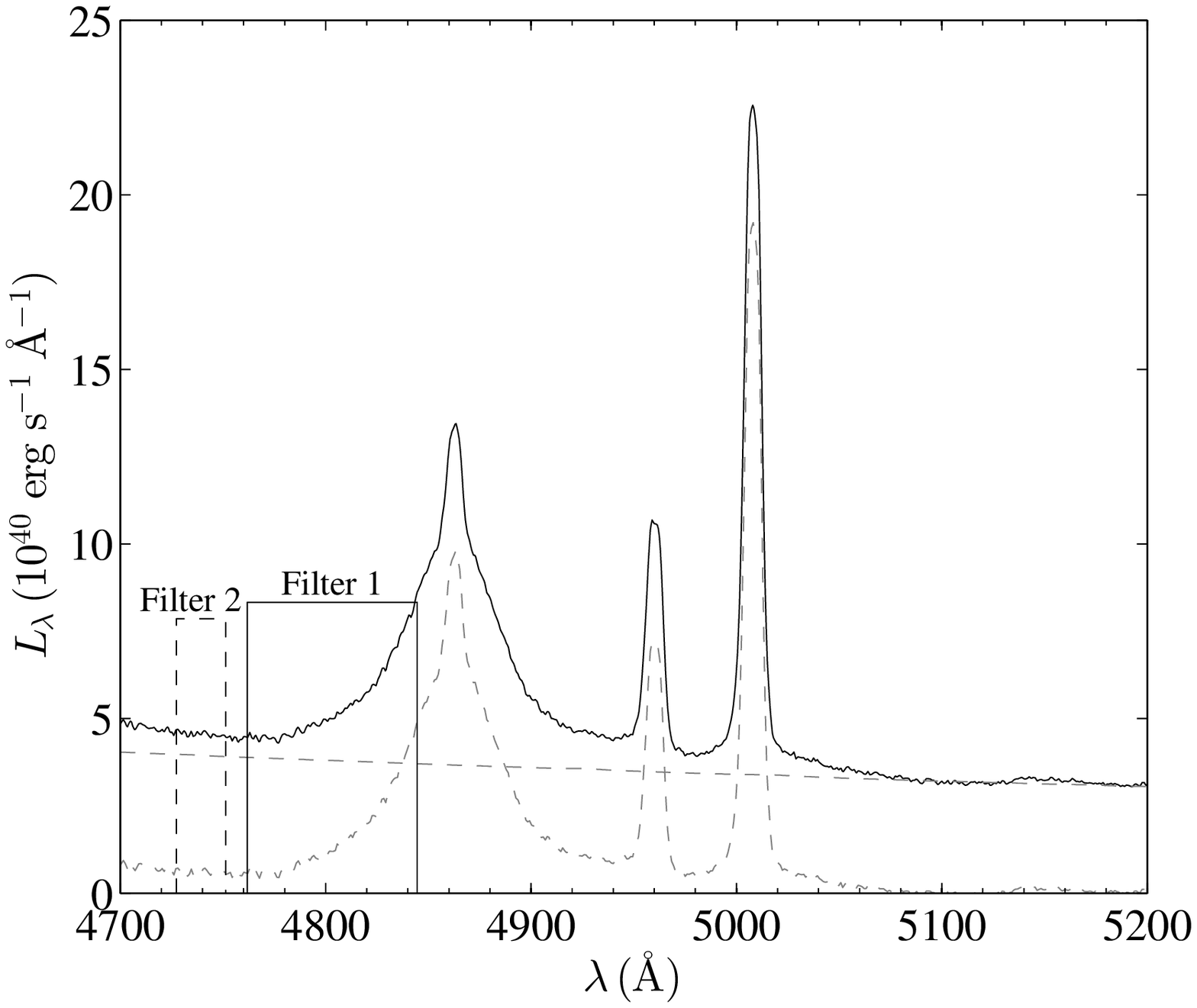}
\caption{{\it Top}: RM masses versus optimal-filter SE masses. The diagonal line is the 1:1 relation. {\it Bottom}: Best-fit positions of the 2 filters superposed on the spectrum of Mrk 509. The solid curve is the original spectrum, and the dashed curves are the decomposed pseudo-continuum and line spectra, respectively. Filter 1 has a positive coefficient and Filter 2 has a negative coefficient, with heights proportional to the absolute value of the coefficient. 
\label{fig:filt2}}
\end{figure}

With 2 filters and all 44 objects in the calibration set, we derive $\sigma_{\rm int}=0.28$ from the best-fit result. We note that the true intrinsic scatter, when the optimal filters are used for prediction, should be somewhat larger than this but smaller than the median $\sigma_{\rm int}$ shown in Figure~\ref{fig:rand}, which is 0.38, because a larger calibration sample is used here. From now on we use this set of optimal filters as our fiducial calibration, and quote a nominal intrinsic mass estimate precision of 0.28~dex. The best-fit parameters of the two filters and the constant coefficients are listed in Table~\ref{tab:filt}. The RM mass versus the estimated SE mass is shown in the top panel of Figure~\ref{fig:filt2}. In the bottom panel of Figure~\ref{fig:filt2}, the locations of the two filters are shown superposed on an AGN (Mrk 509) spectrum from the \citet{Marziani2003} atlas. 

\begin{figure}[ht]
\centering
\includegraphics[width=\columnwidth]{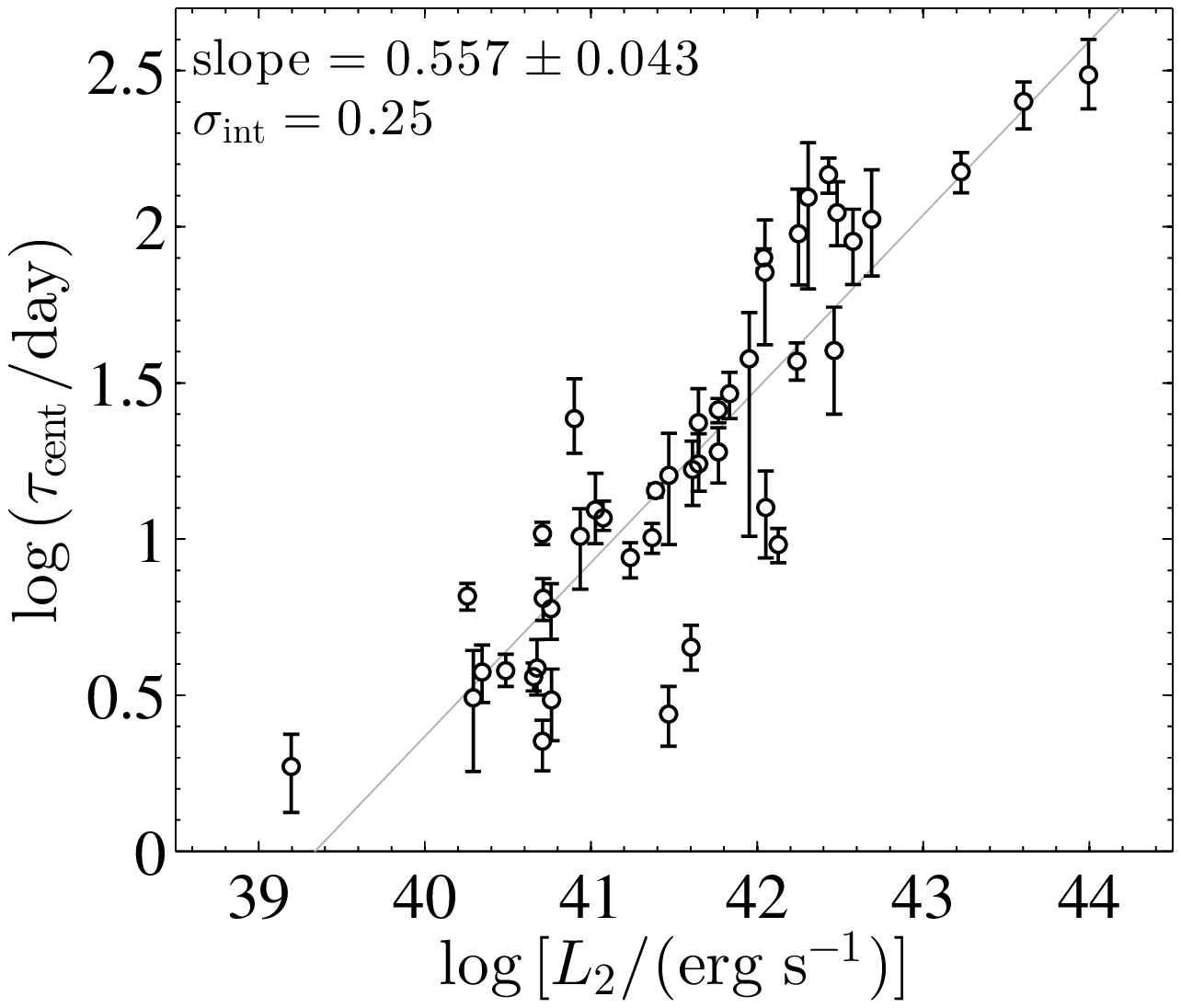}\\
\includegraphics[width=\columnwidth]{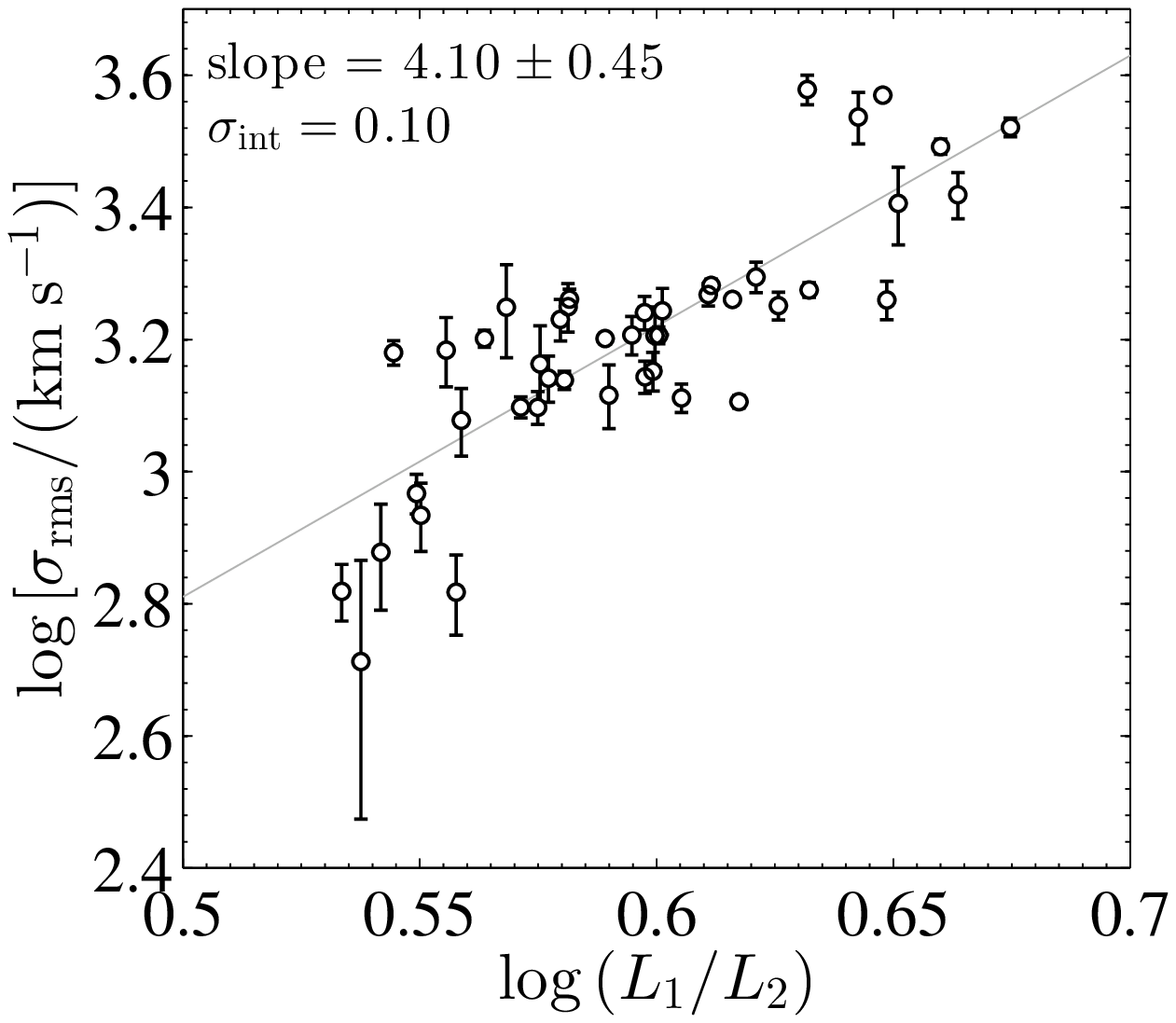}
\caption{$\log \tau_{\rm cent}$ versus $\log L_2$ and $\log \sigma_{\rm rms}$ versus $\log (L_1 / L_2)$ along with the IMVWLS regression result. The slope of the regression line and the intrinsic scatter are shown in the plot, respectively for the two relations. These correlations indicate that the two filters extract virial information embedded in the the spectra needed for black hole mass estimate. 
\label{fig:filtinfo}}
\end{figure}

For the two coefficients associated with the two filtered luminosities, one is positive and the other is negative and their sum $b_1 + b_2 = 0.56$ is very close the to slope of the $R-L$ relation (see Equation~\ref{eq:rl}). The estimated mass using the two filtered luminosities is $M_{\rm SE} \propto L_1^{b_1}L_2^{b_2} = (L_1 / L_2)^{b_1} L_2^{b_1 + b_2}$. If the second term $L_2^{b_1 + b2}$ is reflecting the $R-L$ relation, then the first term $(L_1 / L_2)^{b_1}$ should contain the line width information in the RM mass, i.e., $\sigma_{\rm rms}^2$. In Figure~\ref{fig:filtinfo}, we show that there are indeed correlations, between $\log \tau_{\rm cent}$ and $\log L_2$, and between $\log \sigma_{\rm rms}$ and $\log (L_1 / L_2)$,  respectively.  The intrinsic scatters of the two relations after IMVWLS regression are 0.25 dex and 0.10 dex, respectively, which are as small as those for the $R-L$ and $\sigma_{\rm rms}-$FWHM relations (Equations \ref{eq:rl} \& \ref{eq:fwhm}). Thus, the filters do extract virial information embedded in the spectra.

To further test the validity of the new method, we randomly permute the RM mass for all objects, or randomly assign a mass in the range of $10^6 - 10^9 M_\sun$ for each, and compare the revised RM masses and optimal-filter SE masses. The best-fit $\sigma_{\rm int}$, which was 0.28 dex, becomes 0.8--0.9 dex. This test demonstrates that the filter optimization is driven by physical correlations in luminosity, BH mass and spectral shape, rather than pure mathematical outcomes. 

As the filter-based SE mass estimator is a function of luminosities, one may concern how robust the result will be if biases exist in the measured luminosity density spectrum. In general there are two possibilities: (1) the scale of the luminosity density could be biased by a factor of $f_1$ (e.g., due to an uncertainty in distance or flux calibration), and (2) the AGN spectrum could be contaminated by host galaxy emission by a factor of $f_2$. We find that the $\Delta\log M_{\rm SE} = 0.1$ if $f_1 = 1.50$ or 0.66, suggesting that an error of 20\% in distance (typical for nearby galaxies) results in 0.1~dex in the estimated mass. The same amount of effect is also expected in traditional SE estimators, where $\Delta\log M_{\rm SE} \approx 0.5 \Delta L_{5100}$.The spectrum of the host galaxy consists of a continuum component and a variety of line features. In our case, the continuum is the dominant source of contamination because the filters are wide. For the second test, we increase the pseudo-continuum by a factor of $f_2$ to see how the results vary, and find that $\Delta\log M_{\rm SE} = 0.03-0.04$ if $f_2 = \pm0.1$, or $\Delta\log M_{\rm SE} = 0.05-0.08$ if $f_2 = \pm0.2$.  Adding or subtracting a flat continuum component produces similar results. Usually, the host contamination will not be larger than 20\% of the AGN continuum flux \citep{Bentz2009a,Bentz2013}. The filter-based estimator is slightly less sensitive to host contamination than traditional estimators, where $\Delta\log M_{\rm SE} \approx 0.5f_2$. Therefore, we conclude that the filter-based SE mass estimator is reasonably robust against biases in the luminosity measurements.

\section{Tests and comparisons}
\label{sec:test}

So far, we have constructed three recipes for SE mass estimates:

\begin{enumerate}[A)]

\item Based on naively assumed (theoretical) slopes,
\begin{equation}
\log\left[\frac{M_{\rm SE}}{M_\sun}\right] = 
  \log\left\{
  \left[\frac{L_{5100}}{\rm 10^{44} \; erg\; s^{-1}}\right]^{0.5}
  \left[\frac{{\rm FWHM}}{\rm km\; s^{-1}}\right]^{2}
  \right\} + 0.613
\label{eq:se_vp06}
\end{equation}
with an intrinsic scatter of 0.41~dex. This is the same as in \citet{Vestergaard2006} except for a new zeropoint (was 0.907 in this form of formula). Hereafter we call it the ``updated VP06'' recipe. If one uses host light corrected $L_{5100}$ following Equation~(\ref{eq:hostcorr}), the zeropoint becomes 0.733 and the intrinsic scatter is 0.39.

\item Based on best-fit slopes,
\begin{equation}
\log\left[\frac{M_{\rm SE}}{M_\sun}\right] = 
  \log\left\{
  \left[\frac{L_{5100}}{\rm 10^{44} \; erg\; s^{-1}}\right]^{0.572}
  \left[\frac{{\rm FWHM}}{\rm km\; s^{-1}}\right]^{1.200}
  \right\} + 3.495
\label{eq:se_fwhm}
\end{equation}
with an intrinsic scatter of 0.36~dex. Hereafter $M_{\rm SE}(L_{5100}, {\rm FWHM})$ or FWHM-based SE mass specifically refers to this recipe. With host light corrected $L_{5100}$ (Equation~\ref{eq:hostcorr}), the recipe becomes
\begin{equation}
\log\left[\frac{M_{\rm SE}}{M_\sun}\right] = 
  \log\left\{
  \left[\frac{L_{5100}}{\rm 10^{44} \; erg\; s^{-1}}\right]^{0.504}
  \left[\frac{{\rm FWHM}}{\rm km\; s^{-1}}\right]^{1.200}
  \right\} + 3.602
\label{eq:se_fwhm}
\end{equation}
with an intrinsic scatter of 0.35~dex.

\item Based on two rectangular filters,
\begin{equation}
\log\left[\frac{M_{\rm SE}}{M_\sun}\right] = 
  \log\left\{
  \left[\frac{L_1}{\rm erg\; s^{-1}}\right]^{9.89}
  \left[\frac{L_2}{\rm erg\; s^{-1}}\right]^{-9.33}
  \right\} - 21.55,
\label{eq:se_filt}
\end{equation}
with an intrinsic scatter of 0.28~dex, where $L_1$ and $L_2$ are luminosities filtered from rest-frame wavelength regions listed in Table~\ref{tab:filt}. Hereafter we call it the optimal-filter SE mass and refer to it as $M_{\rm SE}({\rm filter})$.

\end{enumerate}

\begin{figure}[h!]
\centering
\includegraphics[width=0.8\columnwidth]{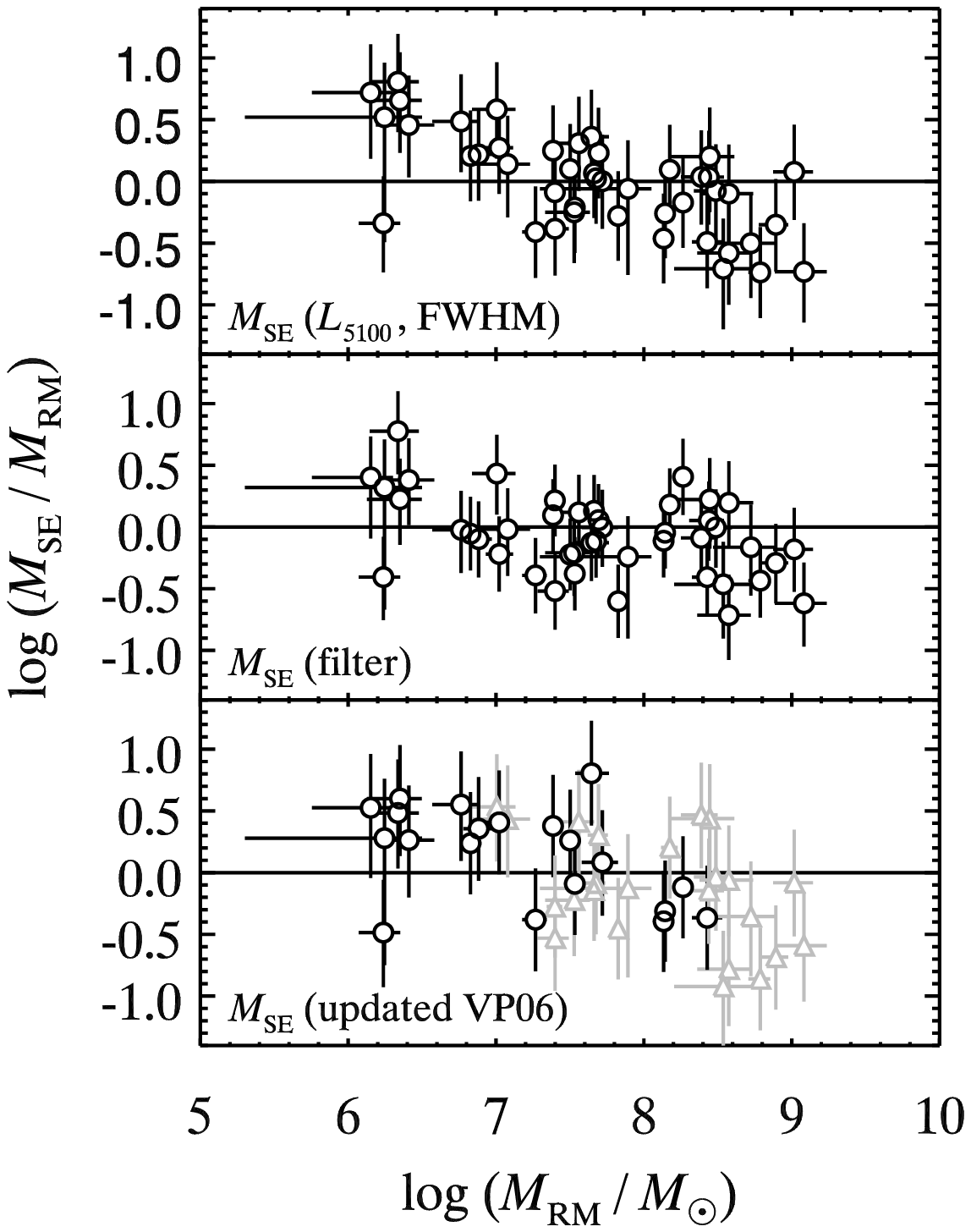}\\
\includegraphics[width=0.8\columnwidth]{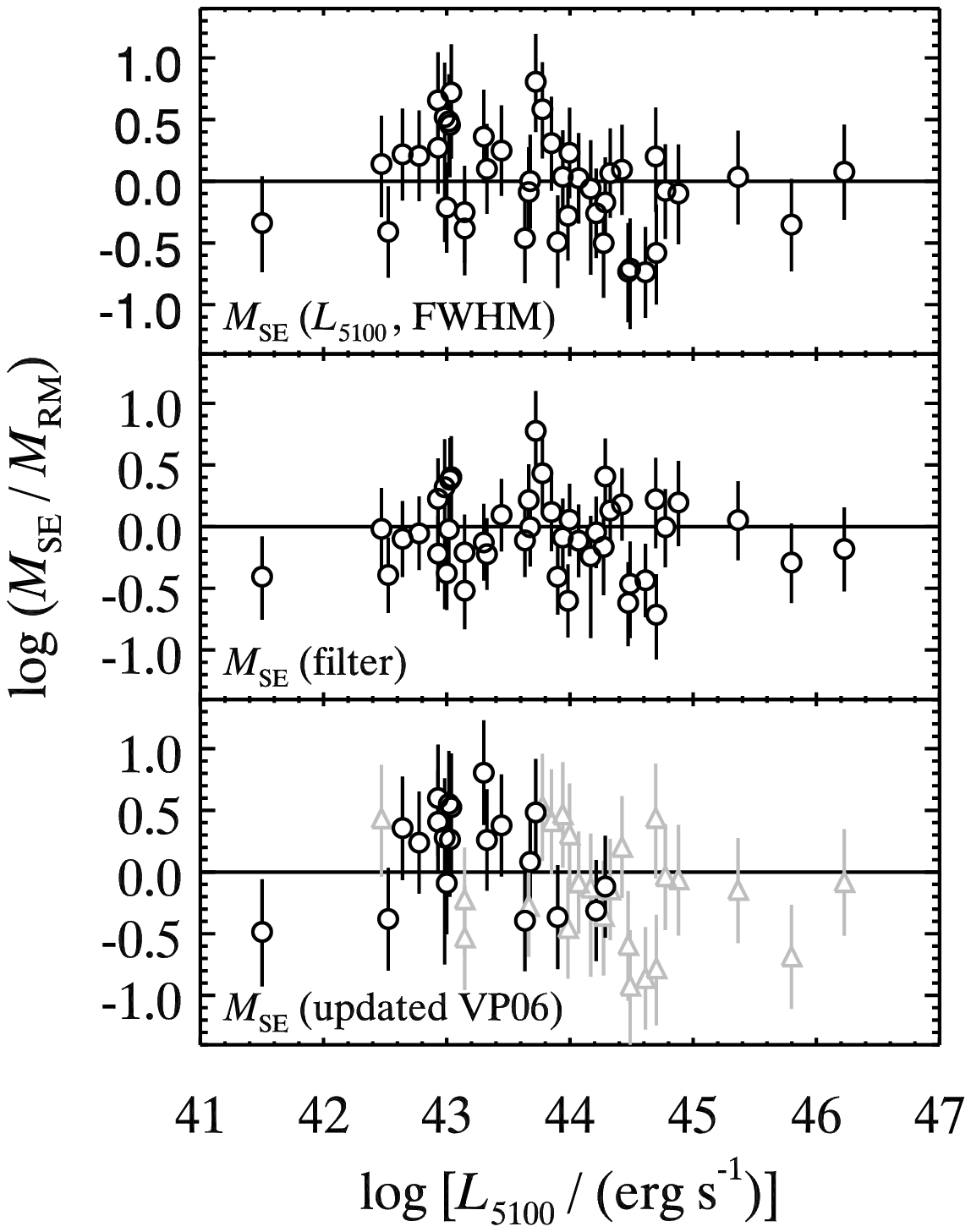}
\caption{The differences between various SE masses and RM masses, as functions of RM mass and SE luminosity. The gray triangles in the bottom panels are the objects included in the VP06 study. 
\label{fig:se_rm}}
\end{figure}

\begin{figure*}[t]
\centering
\includegraphics[width=0.49\textwidth]{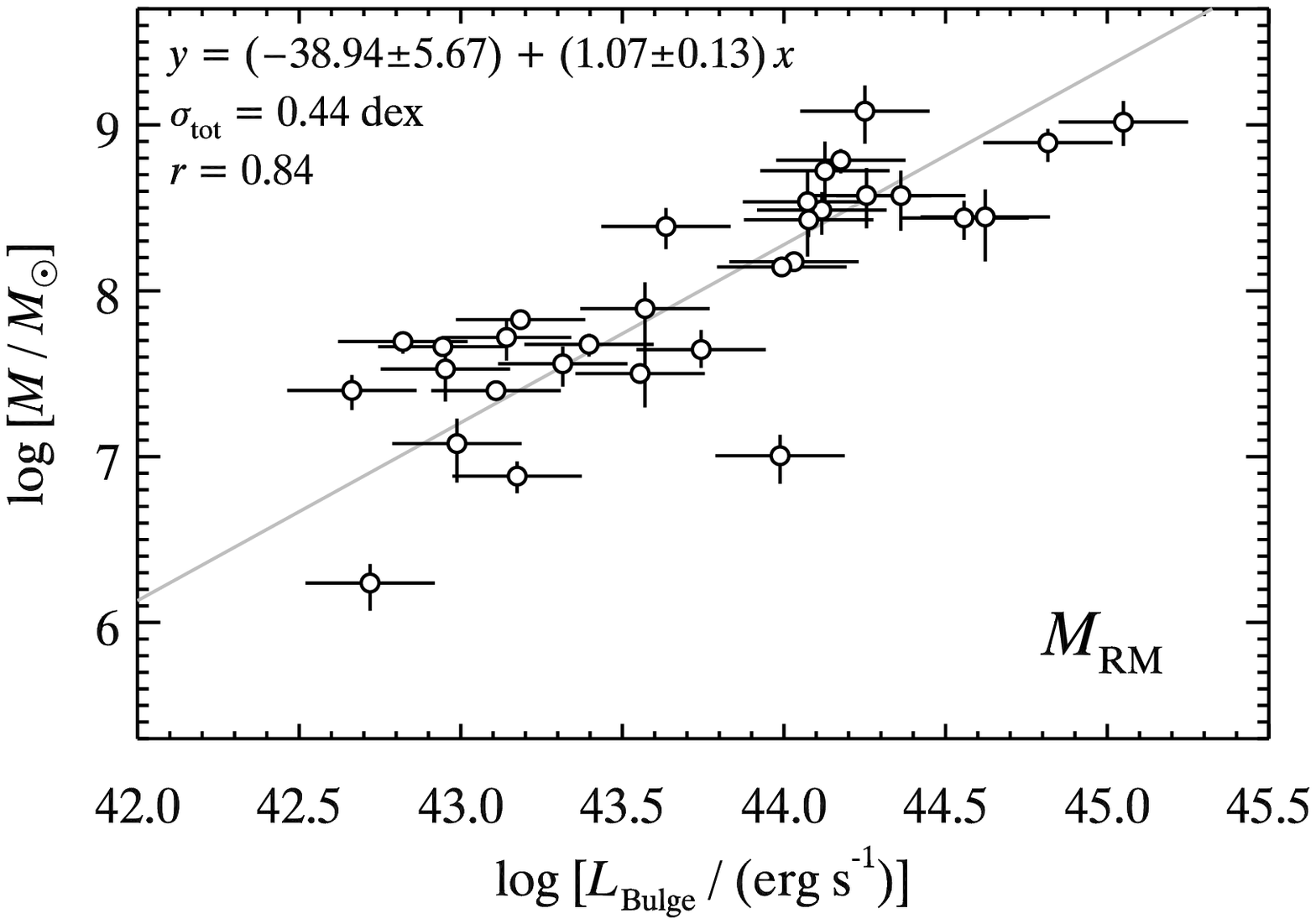}
\includegraphics[width=0.49\textwidth]{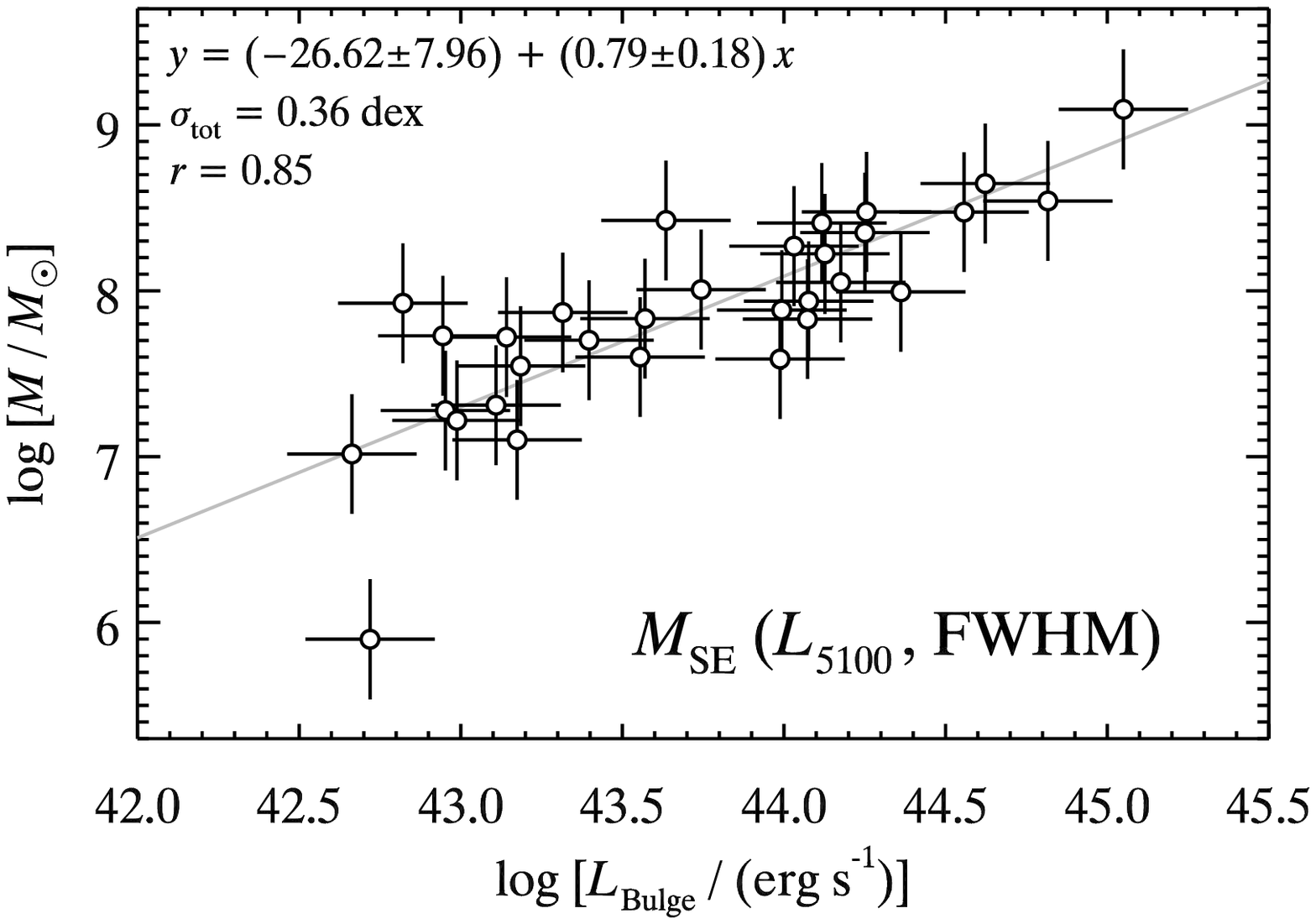}\\
\includegraphics[width=0.49\textwidth]{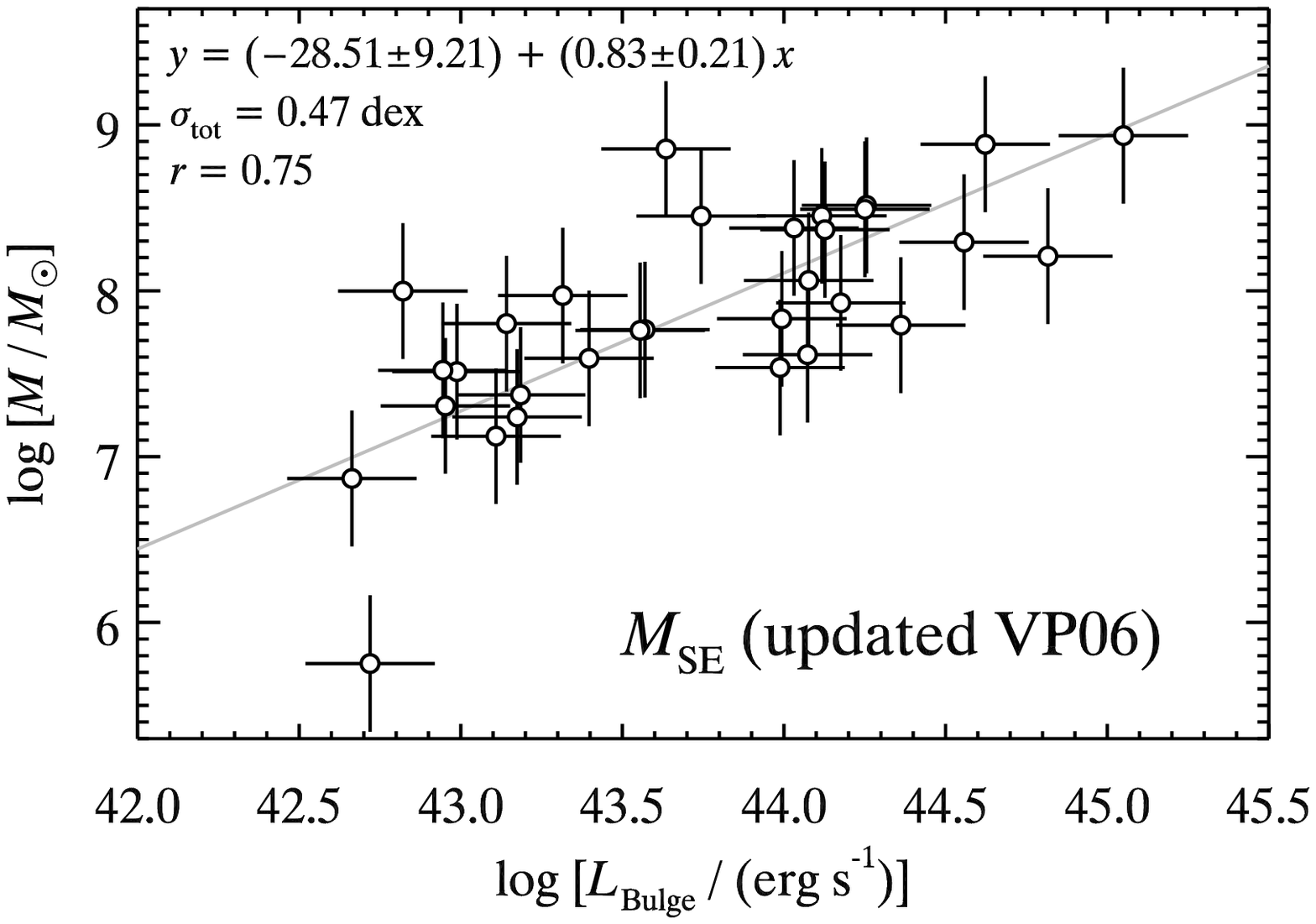}
\includegraphics[width=0.49\textwidth]{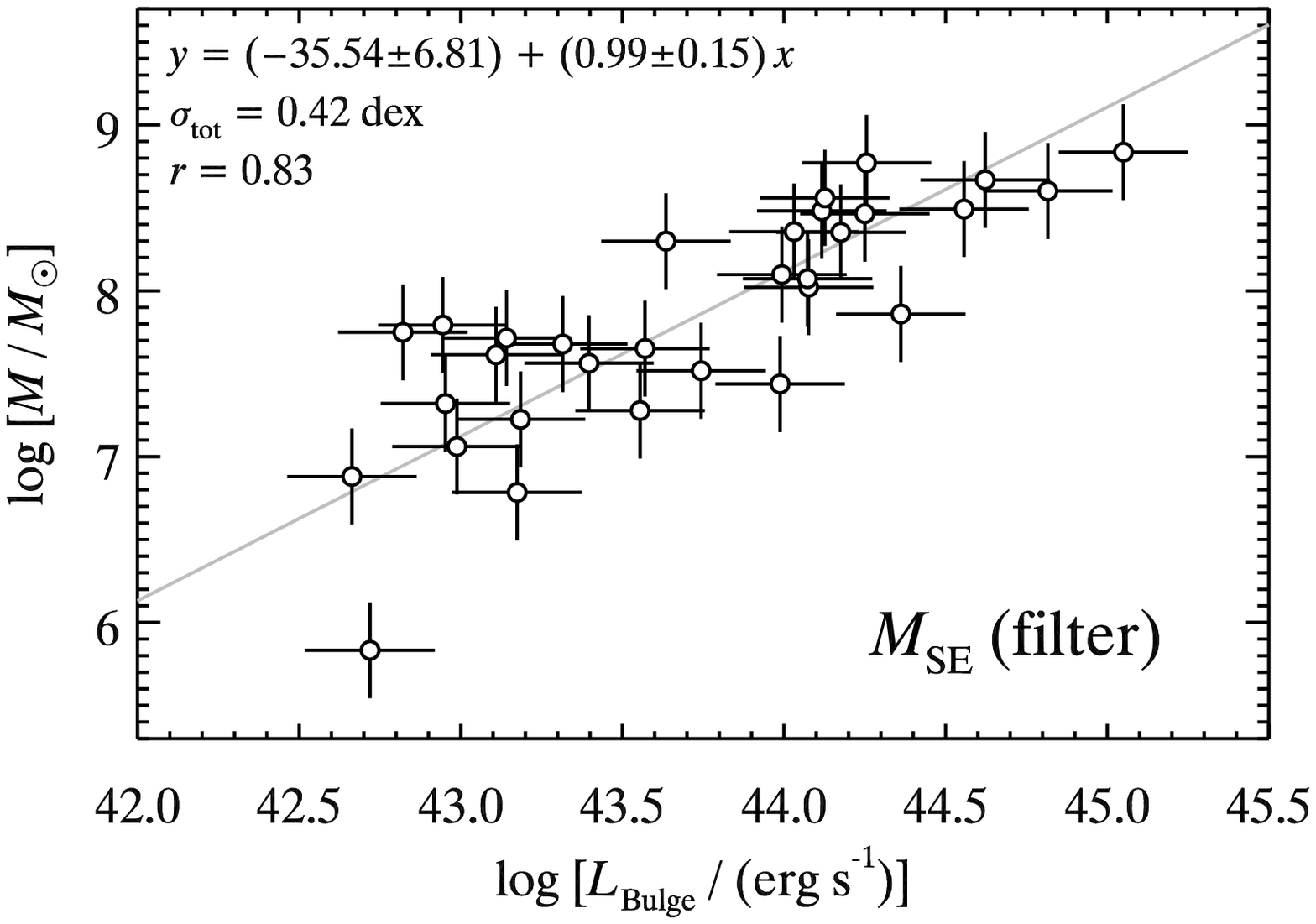}
\caption{Black hole mass versus bulge luminosity for the 31 object in our RM sample with host information. The masses are the RM masses, FWHM based SE masses, updated VP06 masses, and filter based SE masses, respectively. The total scatters after BCES(bisector) linear regression (solid line), and the Spearman's rank correlation coefficients are shown in each panel.
\label{fig:bulge}}
\end{figure*}

For FWHM-based and updated VP06 masses, the error for $M_{\rm SE}$ is quoted as the quadrature sum of the measurement error (in luminosity and line width) and the intrinsic scatter. For $M_{\rm SE}$(filter), it is the quadrature sum of the measurement error (in filtered luminosities) and the semi-amplitude of the 68.3\% prediction interval (see Appendix~\ref{sec:appc}). In most cases, the intrinsic scatter or the prediction interval of the mass recipes dominates the error budget. 
As mentioned above, correcting the host light contribution will change the slope of the $R-L$ relation significantly. However, its effect on the SE masses is negligible (see the last paragraph of \S~\ref{sec:hostcorr}) and the host light correction cannot be done directly for many of the observations in this study. Thus, we will use the recipes without host light correction in the following.

We plot the differences between the SE masses and RM masses as functions of $M_{\rm RM}$ and $\log L_{5100}$ in Figure~\ref{fig:se_rm} to investigate any systematic trend. All SE masses show a residual correlation with $M_{\rm RM}$, indicating an imperfect correlation between SE mass and RM mass. This is simply due to the way of regression. To have an unbiased estimate at fixed $M_{\rm RM}$, one should do regression with the type $(X|Y)$ ($=M_{\rm SE}|M_{\rm RM}$). For FWHM-based SE masses, no regression is done directly against $M_{\rm RM}$ and the regression on individual components are in fact of the type $(Y|X)$ (to predict $Y$ with $X$). For filter-based SE masses, the $(X|Y)$ type is no longer available in the case of multiple linear regression (MLR). However, there is no systematic trend in the mass residuals with SE luminosity. 

The correlation between the black hole mass and the bulge luminosity of the host galaxy can be used to test the accuracy of the SE black hole mass recipes presented here. 31 objects in our sample have bulge luminosities measured at 5100\AA\ with HST imaging data \citep{Bentz2009a}, from which the bulge luminosities are adopted and corrected using our own distances, and assumed to have an error of $\pm0.2$ dex following \citet{Bentz2009b}. The correlation between $M_{\rm RM}$ or different SE masses and bulge luminosities is shown in Figure~\ref{fig:bulge}, along with the regression lines found by BCES(bisector). The total scatters after regression are 0.44~dex, 0.36~dex, 0.42~dex, and 0.47~dex, respectively, for $M_{\rm RM}$, and FWHM-based, filter-based, and the updated VP06 SE masses.  We do not quote the intrinsic scatter because artificial measurement errors for $L_{\rm Bulge}$ are used. The Spearman's rank correlation coefficients are 0.84, 0.85, 0.83, and 0.75, respectively, in the same order. As all these SE masses are calibrated against RM masses, the smaller scatter when using $M_{\rm SE}(L_{5100}, {\rm FWHM})$ is just a coincidence. Taking the slope of the $M_{\rm RM} - L_{\rm Bulge}$ relation as a standard, the filtered SE masses seem to be the most accurate estimator while the FWHM-based SE masses is slightly biased, although the difference in slope is small and within 2$\sigma$ errors.

\begin{figure}[t]
\centering
\includegraphics[width=\columnwidth]{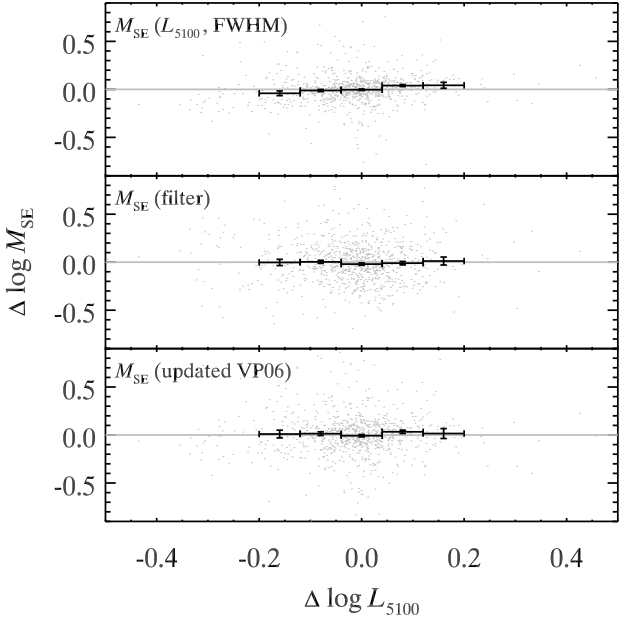}
\caption{Changes in SE mass estimates as a function of changes in luminosity for SDSS quasars with repeated spectroscopy, for the three different recipes.  The bars are medians and standard errors of the dots in each luminosity change interval.
\label{fig:dm_d5100}}
\end{figure}

SDSS quasars with at least two spectroscopic observations separated by 10 or more days are used to test the self-consistency of each SE mass estimator. For objects with more than two spectroscopic epochs available, the earliest and latest were selected. In Figure~\ref{fig:dm_d5100}, we plot the changes in SE masses against the changes in luminosity between the two epochs. Since the BH mass is constant when luminosity varies, we should on average see zero change in the mass estimates when luminosity varies. On the other hand, a positive trend between the changes in SE masses and luminosity will be present if the SE mass recipe does not cope with the luminosity changes, leading to a luminosity-dependent bias \citep[see extensive discussions in e.g.,][]{Shen2013}. Figure~\ref{fig:dm_d5100} shows that the SE recipe based on FWHM is slightly subject to this luminosity-dependent bias, while the filter-based SE and the updated VP06 recipes are not.

\begin{figure}[t]
\centering
\includegraphics[width=0.8\columnwidth]{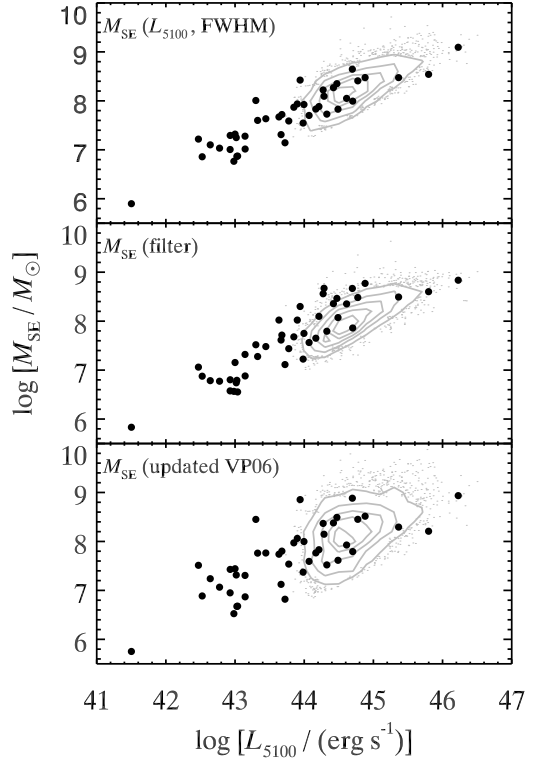}
\caption{Comparison of SE masses for RM AGN (filled circles) and SDSS quasars (gray contours and dots) as a function of AGN luminosity, using the three different recipes. 
\label{fig:ml}}
\end{figure}

\begin{figure}[t!]
\centering
\includegraphics[width=\columnwidth]{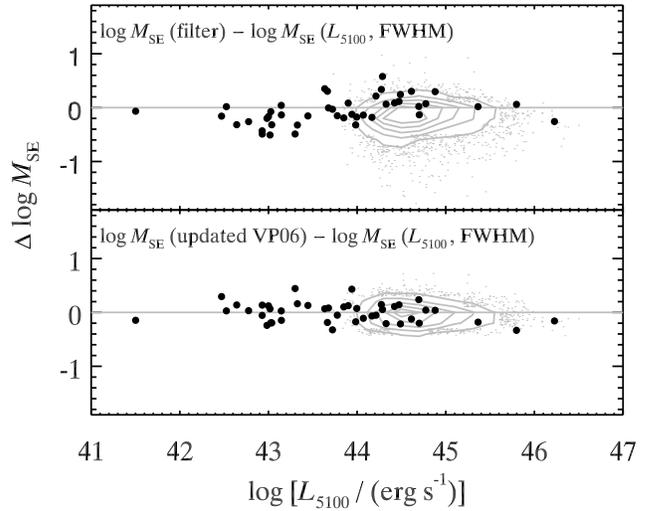}
\caption{Comparison of different SE recipes for RM AGN (filled circles) and SDSS quasars (gray contours and dots) as a function of AGN luminosity. 
\label{fig:dm}}
\end{figure}

Finally, different SE mass estimates are compared using SE spectra from the SDSS DR7 quasar catalog with \hbeta\ coverage and a median $S/N > 20$ in the \hbeta\ region \citep{Shen2011}. The SE masses versus continuum luminosity is shown in Figure~\ref{fig:ml} for both RM AGN and SDSS quasars. The differences between the filter-based or updated VP06 masses and the FWHM-based masses as a function of AGN luminosity are shown in Figure~\ref{fig:dm}. Taking FWHM-based masses for SDSS quasars as the baseline, the filter-based mass is systematically lower by 0.15 dex, and the updated VP06 mass is lower by 0.05 dex, defined by the peak location of the distribution in Figure~\ref{fig:dm}. These systematic differences are significant, reflecting inherent caveats in one or more of these recipes; but they are well below the nominal uncertainties of these mass recipes.

\section{Discussion}
\label{sec:dis}

In this paper we presented three recipes for SE mass estimation, using calibrations against an \hbeta-based RM sample of 44 AGN: an updated VP06 recipe (Equation~\ref{eq:se_vp06}), a FWHM-based recipe adopting best-fit slopes for the two fundamental relations that link SE and RM masses (Equation~\ref{eq:se_fwhm}), and a new, filter-based recipe (Equation~\ref{eq:se_filt}). 

For the updated VP06 recipe, our new zeropoint (=0.613) is 0.29 dex lower than that of VP06 (=0.907), which are likely affected by three factors. First, the VP06 sample may be systematically biased with respect to our sample, see Figure~\ref{fig:se_rm} that most VP06 members in our sample appear to have systematically lower SE virial products, leading to a higher zeropoint to match the RM masses. The zeropoint found by the 24 objects included in VP06 is 0.734. Second, one of the objects PG~2130+099 has a new RM mass \citep[$(46 \pm 4) \times 10^6$~$M_\sun$;][]{Grier2012} that is 10 times lower than that adopted by VP06 \citep[$(457 \pm 55) \times 10^6$~$M_\sun$;][]{Peterson2004} due to new measurements with better sampled lightcurves. The zeropoint given by the 24 objects within VP06 becomes 0.787 if we adopt the old, large RM mass. Finally, the zeropoint is a weighted mean of the difference between RM masses and SE virial products. We include both intrinsic and measurement errors to build the weights, following the idea of IMVWLS, while in VP06 only measurement errors are used. The zeropoint changes from 0.787 to 0.868 if we omit the intrinsic scatter, very close to the VP06 result. This is simply because the fluctuation around the mean is dominated by intrinsic scatters rather than measurement uncertainties. Without considering the intrinsic scatter will give high priority to a few data points that have small error bars but may be remarkably offset from the true center due to the presence of intrinsic scatter. 

However, the updated VP06 estimator has a large intrinsic scatter, 0.41~dex with respect to RM masses. An independent test using the correlation between black hole masses and bulge luminosities also suggests that the updated VP06 estimator is the least precise among all three recipes (Figure~\ref{fig:bulge}). This is because that the SE FWHM \citep[or even the mean FWHM;][]{Collin2006,Wang2009} is not linearly correlated with $\sigma_{\rm rms}$ (Figure~\ref{fig:linewidth}), which is arguably the best indicator of the BLR virial velocity. Thus, the SE mass estimator using best-fit slopes rather than simple assumed slopes for the two fundamental relations ($R-L$ and $\sigma_{\rm rms}-{\rm FWHM}$) gives smaller scatter with respect to RM masses and tighter correlation with bulge luminosities. Despite all these benefits, we want to caution that the SE masses derived using the recipe with best-fit slopes seem to be slightly offset (nearly 3$\sigma$) from a linear correlation with RM masses, see the BCES(bisector) regression in Figure~\ref{fig:reg} and Table~\ref{tab:reg}. Such a bias is mainly due to the choice of a small $\beta$, a shallower slope on FWHM. As mentioned above, the correlation between $\sigma_{\rm rms}$ and FWHM is dominated by a cluster of points in the median region (Figure~\ref{fig:linewidth}). Therefore, $\beta = 1.2$ is so far the best determination of the slope but one may find a better $\sigma_{\rm rms}-{\rm FWHM}$ relation (even not necessarily a power-law relation) in the future once a significantly large RM sample exists.

The filter-based mass estimator has several advantages over the traditional estimators. Most importantly, this method is easy to use and produces robust results even at low $S/N$ because it extracts information from the total flux spectrum directly and no spectral decomposition is needed. Second, it is tested to be precise, with an intrinsic scatter of only 0.28 dex which is the smallest among all known SE estimators. Also, the filter-based mass estimator seems to be the least biased among the three recipes. The test using the correlation between the black hole mass and bulge luminosity (Figure~\ref{fig:bulge}) suggests that the filtered masses are most closely following the correlation seen between RM masses and bulge luminosities (i.e., having a consistent slope). The updated VP06 and the FWHM-based recipes have an offset less than 2$\sigma$. Although the filter-based estimator is a function of luminosities, it shows no correlation with luminosity or the luminosity change (Figures~\ref{fig:se_rm} \& \ref{fig:dm_d5100}), consistent with the virial behavior, while the FWHM-based mass is slightly dependent on the change of luminosity (Figure~\ref{fig:dm_d5100}). To summarize, marginal evidence suggests that the filter-based recipe and the VP06 recipe are slightly more accurate than the FWHM-based recipe. Although found by optimization techniques, the two filters do extract the physics behind the BH mass (Figure~\ref{fig:filtinfo}), i.e., the virial information, in an easier and more robust manner.

This new technique also has great potential in future application. It may not rely on RM masses and any other mass measurements available for AGN can be used as a calibrator. This may allow the filter-based estimator to be calibrated beyond the parameter space occupied by AGN in the RM sample, expanding its scope of application. When the RM sample becomes large enough, this method may be improved by adding more filters to account for subtle features in the spectrum, such as the contamination by the host galaxy and effects due to different AGN types. It will not be a surprise if one finds another set of filters that have completely different positions and widths, because, for instance, luminosities at a variety of wavelengths may be scaled with the BLR size. 

\begin{figure}[t]
\centering
\includegraphics[width=\columnwidth]{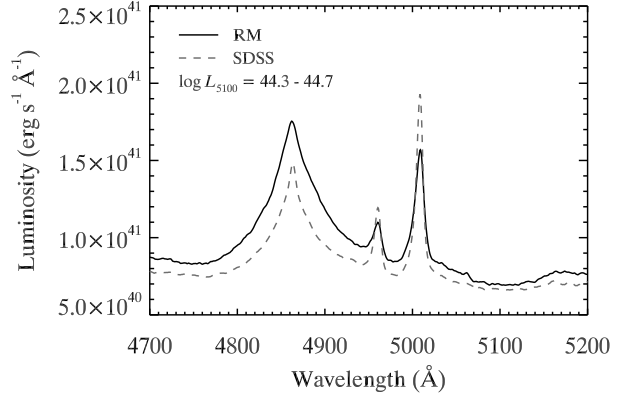}
\caption{Composite spectra of RM and SDSS objects in the $L_{5100}$ luminosity range of $10^{44} - 10^{45}$~\ergs. 
\label{fig:composite}}
\end{figure}

Despite the advantages of this new optimal filter recipe for SE mass estimation, we caution on its inherent limitation and caveats, just as the other traditional SE mass estimators have. One of the main purposes of SE mass estimators is to apply it to large spectroscopic samples of quasars such as those from SDSS. However, all the SE estimators are calibrated using local RM AGN, which may be intrinsically different from distant quasars even with their luminosities matched. Whether or not an estimator can be applied to other quasar samples depends on how the properties of the RM AGN used in the calibration differ from those in other objects. In Figure~\ref{fig:ml}, we compared SE masses for objects in the RM and SDSS samples. For the FWHM-based SE masses, the SDSS quasars seem to be a smooth extension of RM AGN in the $M_{\rm SE}$ vs.\ $L_{5100}$ diagram. The distance from the distribution peak of SDSS objects to the regression center of RM objects is only 0.008~dex in that figure. For the updated VP06 recipe, the deviation is 0.014 dex. However, for the filter-based masses, the RM objects are systematically above the SDSS objects by 0.25 dex in the overlapping luminosity regime. This can be explained by the difference in the spectral shape between RM and SDSS objects. Figure~\ref{fig:composite} displays the composite spectra of RM and SDSS quasars in the luminosity bin of $L_{5100} = 10^{44.3} - 10^{44.7}$~\ergs, a narrow range around the peak of SDSS luminosity distribution. The composite spectra are calculated as the geometric mean of the luminosity density spectra directly, following \citet{VandenBerk2001}, from 6 RM objects and about 1400 SDSS objects, respectively. The two composite spectra show similar broad line width, after removing the continuum component, leading to consistent results in FWHM-based SE masses. However, the filter-based masses have a difference of 0.37~dex, exactly equal to the difference from the peak of the SDSS mass distribution to the median SE mass of the 6 RM objects in this luminosity bin (Figure~\ref{fig:ml} middle). This discrepancy is mainly due to the fact that two composite spectra have a different underlying power-law component.

Due to the lack of an independent check of the mass for SDSS quasars, it is difficult to tell if the SDSS quasars are indeed intrinsically different from local RM AGN with matched luminosity. At least, RM AGN and SDSS quasars do present different spectral shapes as shown in Figure~\ref{fig:composite}. Based on these mass estimates, one may conclude that on average SDSS quasars have a higher Eddington ratio than the local RM AGN, if the filter-based recipe is chosen, or they are emitting at the same Eddington level, if the FWHM-based masses are used. More data are needed to test this, and to assess the limitations of using the RM AGN sample as the sole calibrator.

\acknowledgements 
We thank the anonymous referee for useful comments that have improved the paper, and Aaron Barth and Brad Peterson for kindly providing their calibrated SE spectra for some of the RM AGN included in our analysis. 
HF acknowledges funding support from the National Natural Science Foundation of China under grant No.\ 11222327, the 973 Program of China, and the Tsinghua University Initiative Scientific Research Program.
Support for the work of YS was provided by NASA through Hubble Fellowship grant number HST-HF-51314.01, awarded by the Space Telescope Science Institute, which is operated by the Association of Universities for Research in Astronomy, Inc., for NASA, under contract NAS 5-26555. 
Funding for the SDSS and SDSS-II has been provided by the Alfred P. Sloan Foundation, the Participating Institutions, the National Science Foundation, the U.S. Department of Energy, the National Aeronautics and Space Administration, the Japanese Monbukagakusho, the Max Planck Society, and the Higher Education Funding Council for England. The SDSS Web Site is http://www.sdss.org/.


\appendix

\section{Total and intrinsic variances after BCES regression}
\label{sec:appa}

The bivariate linear regression is to find coefficients $a$ and $b$ to fit a linear relation between two random variables $\xi$ and $\eta$
\begin{equation}
\xi = a + b \eta + \epsilon,
\end{equation}
where $\epsilon$ denotes the intrinsic error. With measurement errors $\sigma_x$ and $\sigma_y$, the measured data are
\begin{equation}
x = \eta + \sigma_x, \quad y = \xi + \sigma_y.
\end{equation}
The residual mean squre gives an unbiased estimation of the total variance after regression,
\begin{equation}
\sigma_{\rm tot}^2 = \frac{\sum_{j=1}^{n}\left( y_j - a - bx_j \right)^2}{n - 2},
\end{equation}
where $n$ is the number of data points (or sample size in this case), and $n - 2$ reflects the degree of freedom (we note that most previous works omitted the degree of freedom when calculating the residual mean square). Assuming $\epsilon$, $\sigma_x$, and $\sigma_y$ follow zero-mean normal distributions, the intrinsic variance for BCES can be derived straightforwardly as
\begin{equation}
\sigma_{\rm int}^2 = \Var(y) - \bracket{ \sigma_y^2 }
  + b^2 \left[ \Var(x) - \bracket{ \sigma_x^2 } \right]
  - 2b \left[ \Cov(x,y) - \bracket{ \sigma_{xy} } \right],
\end{equation}
where Var and Cov are the variance and covariance operators, respectively, and $\sigma_{xy}$ is the covariance of the measurement errors on $x$ and $y$. This form is valid for any type of BCES ($Y|X$, $X|Y$, or bisector), and can be reduced to following forms when specific regression type is used,
\begin{eqnarray}
\sigma_{\rm int}^2 (Y|X) &=& \Var(y) - \bracket{ \sigma_y^2 }
  - b \left[ \Cov(x,y) - \bracket{ \sigma_{xy} } \right], \\
\label{eq:interrxy}
\sigma_{\rm int}^2 (X|Y) &=& b^2\left[ \Var(x) - \bracket{ \sigma_x^2 } \right]
  - b \left[ \Cov(x,y) - \bracket{ \sigma_{xy} } \right].
\end{eqnarray} 

\section{IMVWLS}
\label{sec:appb}

The WLS in \citet{Akritas1996} is different from the transitional one in the sense that the weights include both intrinsic and measurement variances (we call it IMVWLS for distinction). Here we provide an extension of the IMVWLS method from the bivariate case to the more general, multivariate case. Provided $n$ measurements and $k$ independent variables ($k = $ number of filters in our case), the measured data $\mathbf{y}$ and $\mathbf{X}$, measurement errors $\boldsymbol{\sigma}_\mathbf{y}$, and the coefficients $\mathbf{b}$ are denoted in matrix form
\begin{equation}
\mathbf{y} = \begin{bmatrix}
y_1 \\
y_2 \\
\vdots \\
y_n
\end{bmatrix}, \quad
\boldsymbol{\sigma}_\mathbf{y} = \begin{bmatrix}
\sigma_{y,1} \\
\sigma_{y,2} \\
\vdots \\
\sigma_{y,n}
\end{bmatrix}, \quad
\mathbf{X} = \begin{bmatrix}
1 & x_{11} & x_{12} & \cdots & x_{1k} \\
1 & x_{21} & x_{22} & \cdots & x_{2k} \\
\vdots & \vdots & \vdots & \ddots & \vdots \\
1 & x_{n1} & x_{n2} & \cdots & x_{nk} \\
\end{bmatrix}, \quad
\mathbf{b} = \begin{bmatrix}
b_0 \\
b_1 \\
\vdots \\
b_k
\end{bmatrix}.
\end{equation}
The generalized weighted least squares \citep{Montgomery2006} estimator of $\mathbf{b}$ (to fit $\mathbf{y = Xb}$) is 
\begin{equation}
\mathbf{b = \left(X'WX\right)^{-1}X'Wy},
\end{equation} 
where $\mathbf{W}$ is a diagonal matrix consisting of the weight for each observation. We use following steps to do IMVWLS:

\begin{enumerate}[(a)]
\setlength\itemindent{2em}
    \item do ordinary MLR, in which case $\mathbf{W}$ is an identity matrix, $\mathbf{b = \left(X'X\right)^{-1}X'y}$;
    \item calculate the total variance $\sigma_{\rm tot}^2 = \mathbf{(y - Xb)'(y - Xb)} / (n - k - 1)$;
    \item calculate the intrinsic variance $\sigma_{\rm int}^2 = \sigma_{\rm tot}^2 - \boldsymbol{\sigma}_\mathbf{y}'\boldsymbol{\sigma}_\mathbf{y}/n$ (or $=0$ if it is $<0$);
    \item the weight of the $i$th measurement  $w_i = 1 / (\sigma_{\rm int}^2 + \sigma_{y,i}^2)$ and the weight matrix $\mathbf{W} = {\rm diag}(w_1, w_2, \cdots, w_n)$ ;
    \item do IMVWLS, $\mathbf{b = \left(X'WX\right)^{-1}X'Wy}$;
    \item redo (b)--(e); 
    \item the variance of $\mathbf{b}$ is $\mathbf{\boldsymbol{\sigma}_b^2 = {\rm diag}\left[\left(X'WX\right)^{-1}\right]}$.
\end{enumerate}
It is optional to repeat step (f) a few times until $\sigma_{\rm int}$ converges. If the error bars are asymmetric, using the average initially and updating $\boldsymbol{\sigma}_\mathbf{y}$ at each step whenever a new $\mathbf{b}$ is calculated, by choosing the error on the side closer to the regression line.  We note that the total variance for the prediction set is $\sigma_{\rm tot}^2 = \mathbf{(y - Xb)'(y - Xb)} / n$, rather than dividing $(n - k - 1)$ as for the calibration set.

\section{Errors for filter-based masses}
\label{sec:appc}

For prediction, the $100(1-\alpha)$ percent prediction interval \citep[cf.\ Eq.~(3.54) in][]{Montgomery2006} is adopted as the error of filter-based masses. The one-sided error bar size is
\begin{equation}
\sigma = t_{\alpha/2,n-k-1} \sqrt{\sigma_{\rm int}^2 \left( 1 + \mathbf{x_0'(X'X)^{-1}x_0} \right)}, 
\end{equation}
where $t$ denotes the cutoff value of a t-distribution with a probability of $\alpha/2$ and a degree of freedom of $n-k-1 = 41$ ($t = 1.012$ if $1 - \alpha = 68.3\%$), $\mathbf{x_0}' = [1,\; \log L_1,\; \log L_2]$ is the new observable, $\sigma_{\rm int} = 0.28$ is the intrinsic scatter, and $\mathbf{(X'X)^{-1}}$ for our RM sample is 
\begin{equation}
\mathbf{\mathbf{(X'X)^{-1}}} = \begin{bmatrix}
44.7855 &  -1.1504 &    0.0723 \\
0.0723  &  -20.5341 &   20.2422 \\
-1.1504 &   20.8561 &  -20.5341 \\
\end{bmatrix}.
\end{equation}
We note that $\mathbf{(X'X)^{-1}}$ is positive definite and the mininum of $\mathbf{x_0'(X'X)^{-1}x_0}$ is $1/n$ when the prediction is made at the centroid of the sample; the prediction error increases as a function of the distance to the sample centroid.

\end{document}